\documentclass{article}

\usepackage{arxiv}

\usepackage[utf8]{inputenc} % allow utf-8 input
\usepackage[T1]{fontenc}    % use 8-bit T1 fonts
\usepackage{hyperref}       % hyperlinks
\usepackage{url}            % simple URL typesetting
\usepackage{booktabs}       % professional-quality tables
\usepackage{amsfonts}       % blackboard math symbols
\usepackage{nicefrac}       % compact symbols for 1/2, etc.
\usepackage{microtype}      % microtypography
\usepackage{lipsum}		    % Can be removed after putting your text content
\usepackage{graphicx}
\usepackage[square,sort,comma,numbers]{natbib}
\usepackage{doi}
\usepackage{float}
\usepackage{caption}
\usepackage{subcaption}
\usepackage{xcolor}
\usepackage{amsmath}
\usepackage{amsfonts} 
\usepackage{amssymb} 
\usepackage{multirow}
\usepackage{siunitx}

\newcommand{\define}[4]{$#1=#1(#2):#3\mapsto#4$}

\title{A high-order polynomial-corrected shifted boundary method for simulating fully nonlinear water waves} % Using a high-order polynomial-corrected shifted boundary for simulating fully nonlinear water waves with optimal polynomial-preserving gradient recovery

\author{ \href{https://orcid.org/0000-0001-6698-2623}{\includegraphics[scale=0.06]{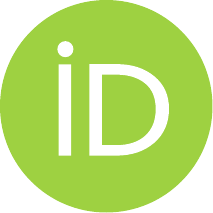}\hspace{1mm}Jens Visbech} \\
	Department of Applied Mathematics and Computer Science\\
	Technical University of Denmark\\
	2800, Kgs. Lyngby \\
	\texttt{jvis@dtu.dk} \\
	\And
	\href{https://orcid.org/0000-0001-8626-1575}{\includegraphics[scale=0.06]{Figures/orcid.pdf}\hspace{1mm}Allan P. Engsig-Karup} \\
	Department of Applied Mathematics and Computer Science\\
	Technical University of Denmark\\
	2800, Kgs. Lyngby \\
	\texttt{apek@dtu.dk} \\
	\And
	\href{https://orcid.org/0000-0002-7263-442X}{\includegraphics[scale=0.06]{Figures/orcid.pdf}\hspace{1mm}Harry B. Bingham} \\
	Department of Civil \& Mechanical Engineering\\
	Technical University of Denmark\\
	2800, Kgs. Lyngby \\
	\texttt{hbbi@dtu.dk} \\
    \And
	\href{https://orcid.org/0000-0002-1679-7339}{\includegraphics[scale=0.06]{Figures/orcid.pdf}\hspace{1mm}Mario Ricchiuto} \\
    INRIA, Univ. Bordeaux, CNRS, Bordeaux INP, IMB, UMR 5251, \\200 Avenue de la Vieille Tour, 33405 Talence cedex, France\\
	\texttt{mario.ricchiuto@inria.fr} \\
}

\renewcommand{\headeright}{Visbech et al. (2026)}
\renewcommand{\shorttitle}{An \textit{arXiv} Preprint}

\hypersetup{
pdftitle={},
pdfsubject={},
pdfauthor={},
pdfkeywords={},
}

\begin{document}

% \input{Outline}
% \newpage

\maketitle

\begin{abstract}
\vspace{-1mm}
	We present a novel unfitted computational framework for simulating fully nonlinear potential flow-based water waves. Focusing on wave propagation, we describe the core methodology, which involves a high-order polynomial-corrected shifted-boundary approximation on unfitted spectral elements. This approach allows for the simulation of a curved, highly time-dependent (moving and deforming) free surface affected by bathymetric changes. All that on a simple Cartesian background mesh without re-meshing or further approximations of non-affine geometric features. In addition, we highlight the importance of proper gradient recovery using a polynomial-preserving technique to accurately capture the vertical free-surface velocity. Ultimately, the goal is to develop a high-order convergent numerical scheme capable of simulating highly nonlinear waves over long periods. This is achieved through an arbitrary-order finite-difference approximation of the free surface combined with added hyperviscosity for numerical stability. 
We present verification and validation test cases for wave propagation in both periodic and finite domains. Emphasis is placed on convergence studies, the justification for using high-order approximations, the importance of optimal gradient recovery, long-time simulation of highly nonlinear waves, and nonlinear wave interactions with both affine and curved bathymetry changes, as well as with vertical walls, for highly nonlinear stream function waves and high-amplitude solitons. 
The novel computational model offers a comprehensive, all-in-one framework for simulating ocean waves and their interactions with offshore structures. It provides significant geometric flexibility, enabling boundaries such as the free surface to move and deform over time without the need to re-mesh a boundary-fitted mesh.
\end{abstract}

% keywords can be removed

% Input your different sections here:
\section{Introduction}\label{sec:introduction}

% Setting the scene: 
As a supplementary method to physical testing in the laboratory or on a full scale in the ocean, numerical models have been used for decades to study wave propagation and wave-structure interaction \cite{evans1987report, tsai1996computation, davidson2020efficient}. Here, circumventing potential scaling issues and the time-consuming process of constructing, modifying, and running costly test campaigns by performing calculations entirely computationally. Consequently, numerical models have become essential tools for the development, design, and analysis of ocean engineering projects, enabling efficient simulations of floating offshore wind turbines, wave energy converters, coastal regions, and much more. However, reliance on simulations imposes significant demands on the numerical model in terms of i) \textit{physical representation} and ii) \textit{numerical accuracy, efficiency, scalability, stability}, and \textit{geometrical flexibility}.

% Requirement i)
Regarding the first requirement, we focus on which governing equations to use, as this sets an upper bound on how close we can be to the real physical truth. Here, the Navier-Stokes (NS) equations are widely recognized as the most precise description of fluid flow, but they also represent the most complex mathematical challenge and demanding computational problem to solve. In many cases, viscous effects are negligible, and a potential flow assumption can be invoked to greatly simplify the mathematical description. Additionally, assumptions and simplifications made to the governing equations and their domain set limits on expectations, such as the exclusion of moving structures or overturning waves. An example of this is the $\sigma$-transformation used in pure wave propagation problems, e.g., see \cite{li1997three, engsigkarup2009efficient}, where the fluid domain is mapped onto a simple, time-invariant domain.

% Requirement ii)
For the second requirement, we are concerned with selecting a numerical method to approximate the mathematical model (the physical representation). We want the numerical method to exhibit convergent error behavior as we refine the approximation resolution, e.g., by refining the mesh in mesh-based methods, increasing the order of the basis functions, or both. This is especially important if low dispersion errors (phase lag) are desired after long-time wave propagation, in which high-order numerical schemes are essential \cite{kreiss1972comparison}. Furthermore, we want the computational cost of the numerical method to scale efficiently with increasing problem size, ideally linearly, such that doubling the number of degrees of freedom doubles the computational work. Equally important, the method should lend itself to efficient parallel computation on modern many-core architectures, for example, through scalable iterative solvers, thereby enabling large-scale simulations over long time periods \cite{engsigkarup2013fast}. The numerical method should also remain stable during temporal integration, ensuring that the solution remains bounded within the physical representation. Finally, applications to real-world problems require substantial geometric flexibility in how the numerical method represents the computational domain and its boundaries. These may be curved and geometrically complex, contain features and curvature spanning multiple length scales, and, in some physical problems, evolve significantly in time through motion, deformation, or even changes in topology.

% Introduce fitted and unfitted methods, highlight different methods, with the end focus on, and introduce the shifted boundary method.
\subsection{Unfitted methods for an efficient, accurate, and flexible geometrical approximation}

% Introduce fitted and unfitted methods:
The modeling of free-surface flows is commonly approached using single-phase or two-phase formulations. In two-phase formulations, both fluids separated by the free surface, such as water and air, are explicitly represented. A widely used approach for capturing the interface in such formulations is the volume-of-fluid (VOF) method, which represents the interface using a volume-fraction field without requiring the computational mesh to conform to the free surface \cite{hirt1981volume}. However, an accurate representation of the interface generally requires coupling the VOF formulation with a sharp-interface technique, either through post-processing or directly within the numerical formulation; see, e.g., \cite{rim2026naturally, cutforth2021efficient} and references therein.

 In this work, we focus on formulations that explicitly track the free surface in a single-phase formulation. To this end, we use the computational setting of embedded/unfitted boundary methods. Classical boundary-fitted methods use meshes that are conformal with all boundaries and interfaces. In contrast, in unfitted methods, the computational domain may be non-conforming with both physical interfaces and domain boundaries. In both approaches, we require a discrete geometrical representation of the domain, i.e., a mesh. For boundary-fitted methods, we have an exact representation (for affine domains using affine elements and for curved domains using curvilinear elements); however, this comes at the cost of re-meshing or mesh updating if the domain is time-dependent. This can be very time-consuming, and as the domain's complexity increases, mesh generation becomes dependent on a sophisticated mesh generator and an experienced engineer to operate it.
On the other hand, for unfitted methods, generally speaking, the approach is to decouple the mesh generation and the domain in two steps: i) generate a very simple Cartesian mesh and ii) immerse/embed the domain onto this mesh. Then, the last step is to incorporate the appropriate geometric representation and the boundary conditions for the chosen mathematical problem. It is here that the various unfitted methods primarily differ.

% Review some of the most classical unfitted methods:
Building on Peskin's 1970s work \cite{peskin1972flow}, the immersed boundary method (IBM) was introduced. The literature on immersed/embedded/unfitted methods is vast, and one can hardly mention all the methods. To name a few, we mention the ghost fluid method (GFM) \cite{fedkiw1999non}, the cut finite element method (CutFEM) \cite{burman2015cutfem}, the aggregated unfitted finite element method (AgFEM) \cite{badia2018aggregated}, and very recently, the unfitted $\phi$ finite element method ($\phi$-FEM) \cite{duprez2023new}. 
Although all of the methods mentioned above avoid the need for boundary-fitted meshes, they differ in how they represent the true geometry and enforce boundary conditions. Earlier approaches, such as IBM and GFM, typically modify the governing equations or add forcing terms to handle immersed boundaries. More recent embedded finite element methods, including CutFEM and AgFEM, aim to perform computations on unfitted meshes while maintaining the accuracy, stability, and flexibility of standard finite element discretizations, even in the presence of cut elements or implicitly defined domains. The \textit{small-cut-cell} stability problem is well known, has led to many stabilization techniques (see \cite{birke2023dod, berger2024new} and references therein), and has inspired some of the formulations mentioned above. Among the methods that allow one to sidestep this issue, we have the $\phi$-FEM and the shifted boundary method (SBM), introduced by Main and Scovazzi \cite{main2018shifted_part1, main2018shifted_part2}, with the latter methodological framework used as the basis in this work.

 In the SBM, instead of cutting the boundary intersecting elements, they are kept whole. The SBM then solves the problem on this artificial/approximate \textit{surrogate} domain rather than on the actual \textit{true} domain. This is done by modifying the boundary condition itself so that the modified condition on the surrogate domain is consistent with a high-order approximation of the true condition on the true boundary. More specifically, this is done using a unique mapping between points on the two boundaries and Taylor series expansions, ultimately consistently satisfying the boundary conditions while preserving optimal convergence. The SBM has matured in recent years, and its literature is vast. We refer to some of its main works \cite{main2018shifted_part1, main2018shifted_part2, song2018shifted, nouveau2019high}, with a high-order version in \cite{atallah2022high} and the weighted SBM (WSBM) in \cite{colomes2021weighted}.
 
%  The SBM has matured in recent years by solving fundamental mathematical problems (elliptic problems, hyperbolic systems, etc.); see, e.g., \cite{main2018shifted_part1, song2018shifted, nouveau2019high, atallah2022high}, as well as problems related to fluid and solid mechanics; see, e.g., \cite{main2018shifted_part2, atallah2021shifted, li2023simple, colomes2021weighted}.
% % The extensions of SBM: 
% Methodological extensions to the original SBM include the extension to high-order \cite{atallah2022high}, the weighted SBM (WSBM) \cite{colomes2021weighted}, the SBM in iso-geometric analysis (SBM-IGA) \cite{antonelli2024shifted}, the Gap-SBM \cite{collins2026gap}, and the generalized SBM (GSBM) \cite{colomes2026generalized}.

% Introduce the FNPF and perform a short literature review on this:
\subsection{Simulating waves using unfitted discretizations}
For this work, we use the fully nonlinear potential flow (FNPF) approximation to simulate water waves \cite{dean1991water}. With this, at each discrete time step/stage, we solve a Laplace problem in the fluid domain, coupled with two nonlinear free-surface boundary conditions that evolve in time. The FNPF formulation arises from the NS equations \cite{engsigkarup2013fast} and is subject to various assumptions about the fluid (incompressible, inviscid, and irrotational), and is regarded as a reasonable physical representation for modeling ocean waves and their interactions with offshore structures. Arguably, when a wave problem is applicable to FNPF and is solved efficiently, it can be multiple orders of magnitude faster than conventional VOF-based CFD simulations \cite{ransley2019blind} at the same level of accuracy. When simulated, the geometrical features of the domain are altered at each discrete time step/stage due to moving, deforming, curved, and complex boundaries (free surface, bathymetry, structures, etc.), ultimately requiring re-meshing/updating at all times.
Moreover, linear potential flow (LPF) is applicable if further simplifications can be made to the problem, such as a small wave height relative to the wavelength and small body motion relative to the body length \cite{newman2018marine}. Additionally, LPF offers additional orders-of-magnitude improvements in performance compared to the FNPF problem. The LPF is a simplified model formulation derived by assuming a small-amplitude wave, which eliminates the need for time-dependent mesh updates and the nonlinear terms in the FNPF formulation. Yet, it still poses challenges in representing the potentially curved and complex geometry of floating or submerged structures, necessitating unfitted discretizations.

% Literature review:
Focusing on the LPF and FNPF formulations, we review some of the advances that have been made for wave modeling on unfitted discretizations, including both pure wave propagation and wave-structure interactions:
% Immersed:
Building on \cite{bingham2007accuracy,engsigkarup2009efficient}, where the finite difference method (FDM) is used to solve the FNPF wave problem, we have the work presented in \cite{kontos2016robust,hicks2020development}. Here, an IBM approach was used to model wave propagation (through a $\sigma$-transform) and wave-structure interactions, e.g., using a wedge-shaped wave-maker to generate nonlinear waves \cite{hicks2021nonlinear}. Later extensions to these works are reported in \cite{xu2021finite,xu2023high} for a finite Eulerian/Lagrangian non-transformed domain. Additionally, \cite{knoblauch2026development} recently presented a similar approach for modeling FNPF wave propagation using the FDM and the IBM. In \cite{tong2024adaptive}, the IBM approach has also been combined with the harmonic polynomial cell (HPC) method, originally proposed in \cite{shao2014harmonic}, to solve FNPF problems for wave–structure interaction.
% Overlapping grids:
A partially unfitted approach is to use an overlapping grid (OG), in which the domain and boundaries have individual boundary-fitted meshes that overlap and are unfitted to each other \cite{chesshire1990composite}. This concept has also been combined with the FDM in \cite{amini2017solving,amini2018pseudo} to compute LPF radiation and diffraction problems, and in \cite{amini2019nonlinear} to present preliminary results on FNPF computations with and without structures. Moreover, solving hydro-elastic problems in \cite{zhou2024solving}.
% Combining IBM and OG:
Combining both IBM and OG methodologies with the HPC method is the work in \cite{hanssen2021potential}, which simulates nonlinear wave-making and surface-piercing wave-structure interactions.

% SBM and free surface flows:
Considering fluid-structure interactions and free-surface flows (not solely in connection with FNPF or LPF) in the context of SBM: The shallow water equations are solved in \cite{song2018shifted} and the incompressible NS equations subject to a free surface are solved in \cite{colomes2021weighted, xu2025weighted} using the WSBM approach. The high-order SBM was used in \cite{atallah2025high} to model interfaces for Lagrangian shock hydrodynamics. Moreover, the incompressible NS equations are modeled using an adaptive Octree-based technique with the SBM in \cite{yang2025octree}. Very recently, the LPF problem of estimating hydrodynamic coefficients (added mass and damping) in the frequency domain was solved in \cite{modderman2025application} using lower-order SBM, CutFEM, and AgFEM. As a follow-up, extensions to the WSBM are given in \cite{modderman2026unfitted}.

% Emphasis on missing FNPF unfitted work
Even though FNPF wave propagation and wave-structure interaction pose highly time-dependent problems (with multiple curved, complex boundaries that move and deform over time), making them well-suited to unfitted domain discretizations, the literature review above indicates that this is a relatively underexplored research topic.

% Classical paper contribution:
\subsection{Paper contribution}

Motivated by this realization, we introduce a new unfitted approach based on the SBM for modeling FNPF-based water waves. The new model presents itself as a high-order unfitted model employing a high-order spectral element-based polynomial-corrected shifted boundary approximation combined with optimal polynomial-preserving gradient recovery. The primary numerical discretization of the fluid domain is handled weakly using unfitted spectral elements to obtain a solution for the Laplace problem. Motivated by stability, the nonlinear free-surface conditions are handled strongly via a collocation scheme using arbitrary-order finite-difference approximations and added numerical hyperviscosity.

We outline the entire methodology step by step, emphasizing the polynomial-corrected shifted boundary approximation, as the application of this unfitted approach is novel in the field of computational models for FNPF wave simulation. Several verification and validation tests are conducted in both periodic and finite wave tanks. These include: i) convergence studies, ii) speed-up measurements to justify using high-order methods instead of traditional low-order ones and underlining the importance of proper gradient recovery of the vertical free surface velocity, iii) long-term wave propagation simulations of highly nonlinear waves with and without wave generation and absorption zones, iv) the generation of high-order harmonics over a submerged bar, and v) nonlinear soliton interactions with a vertical wall and a half-circular bathymetry. This supports the validity of the proposed unfitted high-order wave model based on the methodology in \cite{visbech2025spectral}. Very preliminary results have been reported in \cite{visbech2024high,visbech2026recent}; however, this work extends the initial results by improving gradient recovery, numerical stability, and accuracy, and by introducing new nonlinear test cases. This work can be seen as a stepping stone towards the inclusion of fully or partially submerged structures, ultimately providing a complete unfitted framework for simulating all wave propagation and wave-structure interaction problems using FNPF.

% Classical paper outline:
\subsection{Outline}
Following this section's introduction, we present the continuous mathematical model in Section \ref{sec:math_model}, defining the domain and the governing equations. Hereafter, we outline the numerical discretization of the fluid domain in Section \ref{sec:numerical_dis_fluid_domain} and the free surface in Section \ref{sec:numerical_dis_free_surface}, covering topics such as the construction of the unfitted domain, the polynomial corrected shifted boundary approximation of the Laplacian, the discretization of the free surface, polynomial preserving gradient recovery, and time integration and stabilization. The paper's main numerical results are discussed in sections \ref{sec:results_periodic} and \ref{sec:results_finite}, with additional material reported in the appendix. Finally, Section \ref{sec:conclusion} summarizes the main results of the work and discusses perspectives and future directions. 
\section{Mathematical model}\label{sec:math_model}

\begin{figure}[h]
    \centering
    \includegraphics{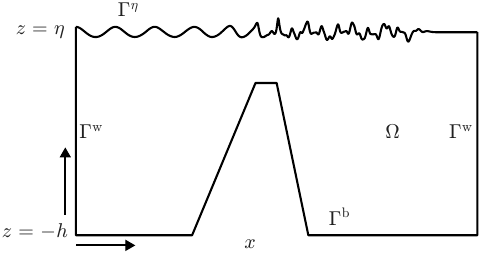}
    \caption{Illustration of the fluid domain, $\Omega$, and its boundaries segments $\Gamma^i$ with $i\in\{\eta,\text{w},\text{b}\}$.}
    \label{fig:fluid_domain}
\end{figure}

We consider a time-dependent, potentially curved, two-dimensional (2D) fluid domain denoted by $\Omega \in \mathbb{R}^2$. The domain is spanned by the horizontal $x$-axis and the vertical $z$-axis, and is completely surrounded by its boundary $\partial \Omega =\Gamma \in \mathbb{R}^2$. For this work, we partition $\Gamma$ into: 
i) the time-dependent (moving and deforming) free surface, $\Gamma^{\eta}$, at $z = \eta$, where \define{\eta}{x,t}{\Gamma^{\eta} \times \mathcal{T}}{\mathbb{R}} is the free surface elevation defined from $z = 0$. Moreover, $\mathcal{T}:t \geq 0$ is the time domain. 
ii) the bathymetry, $\Gamma^{\text{b}}$, at $z = -h$, where \define{h}{x}{\Gamma^{\text{b}}}{\mathbb{R}} is the height of the water column measured from $z = 0$. 
iii) walls, $\Gamma^{\text{w}}$, defined vertically at the ends of $\Omega$. 
iv) periodic boundaries, $\Gamma^{\text{p}}$, also defined vertically at the ends of $\Omega$. With all of this, we have that $\Gamma = \Gamma^{\eta} \cup \Gamma^{\text{b}} \cup \Gamma^{\text{w}} \cup \Gamma^{\text{p}}$ which completely encloses $\Omega$. See Figure \ref{fig:fluid_domain} for a conceptualization of the fluid domain and its boundaries. Structures (partially or fully submerged) can be included in this setting, yet this is not the focus of the present work.

\subsection{Governing equations}

We adopt a potential-flow approximation in two space dimensions, assuming the fluid is incompressible, inviscid, and irrotational. The fluid velocities, $\boldsymbol{u} = (u,w) = (u(x,z,t),w(x,z,t)) : \Omega \times \mathcal{T} \mapsto \mathbb{R}^2$, can then be expressed as the gradient of a velocity potential, $\boldsymbol{u} = \boldsymbol{\nabla} \phi$, where \define{\phi}{x,z,t}{\Omega \times \mathcal{T}}{\mathbb{R}} with $\boldsymbol{\nabla} = (\partial_x, \partial_z)$ being the gradient operator on $\Omega$. The mathematical problem is to find $\phi \in C^2(\Omega)$ by solving the Laplace problem as
\begin{subequations}
\label{eq:Laplace_strong}
\begin{align}
    \boldsymbol{\nabla}^2 \phi  &= 0, \quad \text{in} \quad \Omega, \\
    \phi &= \phi_{\eta}, \quad \text{on} \quad \Gamma^{\eta}, \\
    \boldsymbol{\nabla} \phi \cdot \boldsymbol{n} &= 0, \quad \text{on} \quad \{\Gamma^{\text{b}},\Gamma^{\text{w}} \}, \\
    \phi |_{x = x_{\min}} &= \phi |_{x = x_{\max}}, \quad \text{on} \quad \Gamma^{\text{p}},
\end{align}
\end{subequations}
where the subscript notation of "$\eta$" indicates variables defined in $\Omega$, but also on $\Gamma^{\eta}$, e.g., the free surface velocity potential, \define{\phi_{\eta}}{x,t}{\Gamma^{\eta} \times \mathcal{T}}{\mathbb{R}}. Moreover, $\boldsymbol{n} = (n_x,n_z) = (n_x(x,z,t),n_z(x,z,t)): \Gamma \mapsto \mathbb{R}^2$ is the outwards-facing normal vector to the boundary and any partition hereof. From the solution to the Laplace problem, we can compute the vertical free surface velocity, \define{w_{\eta}}{x,t}{\Gamma^{\eta} \times \mathcal{T}}{\mathbb{R}}, as $w_{\eta} = \partial_z \phi |_{z=\eta}$. This connects to the two coupled time-governing nonlinear free surface conditions, the kinematic and dynamic, on Zakharov form as 
\begin{subequations}
\label{eq:free_surface_equations}
\begin{align}
    \partial_t \eta &= w_{\eta} - \partial_x \eta \partial_x \phi_{\eta} +  w_{\eta} \partial_x \eta \partial_x \eta, \quad \text{on} \quad \Gamma^{\eta} \times \mathcal{T}, \\
    \partial_t \phi_{\eta} &= - g \eta - \frac{1}{2} \partial_x \phi_{\eta} \partial_x \phi_{\eta} +  \frac{1}{2} w_{\eta}^2 (1 +  \partial_x \eta \partial_x \eta), \quad \text{on} \quad \Gamma^{\eta} \times \mathcal{T}, 
\end{align}
\end{subequations}
where $g = 9.81~[\si{m/s^2}]$ is the gravitational acceleration and $\partial_t$ denotes the time-derivative. A linearized version of \eqref{eq:free_surface_equations} can be obtained by dropping all terms but the first on the right-hand side of both equations and applying them on $z=0$.

\subsubsection{Pressure and forces}

Using Bernoulli's equation, we can compute the local pressure, including both hydrostatic and dynamic contributions, at any point in the fluid domain. Integrating this over any wetted boundary, or partition hereof, denoted by $S$, gives the instantaneous force, $\boldsymbol{F} = (F_x,F_z) = (F_x(t),F_z(t)): \mathcal{T} \mapsto \mathbb{R}^2$, as
\begin{equation}\label{eq:force}
    \boldsymbol{F} = - \rho  \int_{S} \left ( gz + \partial_t\phi + \frac{1}{2}\nabla\phi \cdot \nabla \phi \right ) \boldsymbol{n} ~d S, \quad \text{in} \quad \mathcal{T},
\end{equation}
where $\rho = 1,025 ~ [\si{kg/m^3}]$ is the density of water and $\boldsymbol{n}$ is the normal vector to $S$.

\subsection{Wave initialization, generation, and absorption}\label{sec:ini_gen_abs}

In this work, we will consider two main configurations: a periodic wave tank with constant depth; a finite wave tank with the bathymetry changes. For the first case, we initialize our wave simulation with $\eta$ and $\phi_{\eta}$ based on the stream function solution \cite{fenton1988numerical} to the FNPF model. For this, we specify the dimensionless dispersion parameter, $kh$, from which the depth, $h$, is computed, as the domain/wave length is fixed to $L = 1~[\si{m}]$ such that the wave number $k=2\pi/L$ is known. Moreover, we also specify the relative wave steepness, $\varepsilon/\varepsilon_{\max}$, between $0\%$ and $100\%$ to set the nonlinearity of the given wave. Here, $\varepsilon = H/L$ is the actual wave steepness based on the wave height, $H$, and $\varepsilon_{\max}$ is the maximum possible steepness before breaking. Lastly, the wave period is denoted by $T$ and must satisfy the dispersion relation of the waves such that $c=L/T$.

For the finite wave tank, solutions for $\eta$ and $\phi_{\eta}$ are initialized to zero. We also consider solitary waves, which we initialize with the spectrally accurate method presented in \cite{dutykh2014efficient}. This method provides solution values for $\eta$ and $\phi_{\eta}$ on a grid different from what is used in this work; see Sections \ref{sec:SEM} and \ref{sec:finite_difference}. We work around this by computing the soliton solution on a very fine grid and then spline-interpolating to our desired grid spacing, thereby preserving the method's accuracy \cite{dutykh2014efficient}.
Wave generation and absorption are managed through relaxation zones \cite{larsen1983open}. We denote the generation and absorption zones as $ \{ \Gamma^g, \Gamma^a\} \in \Gamma^{\eta}$, and post-process the free surface variables at each timestep as $f =(1-C_r)f + C_rf_{\text{true}}$, with $f \in \{\eta, \phi_{\eta} \}$ and \define{C_r}{x,y}{\Gamma^{\eta}}{\mathbb{R}} is the relaxation shape function. For $C_r$, we employ a high-order step function as $C_r = 6 \zeta^5 - 15 \zeta^4 + 16\zeta^3$, with
\begin{equation}
    \zeta = \frac{|(x-x_0)-L_g|}{L_g}, \quad \text{for} \quad x_0 \leq x \leq x_0 + L_g, \quad \text{and} \quad
    \zeta = \frac{|(x-x_{\text{end}})+L_a|}{L_a}, \quad \text{for} \quad x_{\text{end}} - L_a \leq x \leq x_{\text{end}},
\end{equation}
where $x_0$ and $x_{\text{end}}$ are the beginning and end of the domain, respectively, and $L_g$ and $L_a$ are the length of the generation and absorption zone, respectively. In $\Gamma^g$, $f_{\text{true}}$ is the actual stream function wave solution, \cite{fenton1988numerical}, whereas $f_{\text{true}} = 0$ outside. The wave generation is gradually (linearly) ramped up over five wave periods to minimize the potential impulsive behavior of the system. 
\section{Numerical discretization: The fluid domain}\label{sec:numerical_dis_fluid_domain}

In the following section, we outline the numerical discretization of the fluid domain and the Laplacian in the context of a high-order polynomial corrected shifted boundary approximation on an unfitted computational domain.

\subsection{Construction of the unfitted domain}

We start out by denoting a conformal computational domain by $\Omega_h$ and an unfitted computational domain by $\widetilde{\Omega}_h$. It is possible that $\Omega = \Omega_h$, at least when $\Omega$ has piecewise polynomial boundaries. Conversely, we always have $\Omega \neq \widetilde{\Omega}_h$. From now on, we refer to $\Omega$ as the \textit{true} domain, and to $\widetilde{\Omega}_h$ as the unfitted \textit{surrogate} domain. To construct $\widetilde{\Omega}_h$ we follow the three steps hereafter:
\begin{itemize}
    \item[i)] Define a regular, uniform background tesselation of non-overlapping finite elements denoted by $\mathcal{T}_h$, with $\text{clos}(\Omega) \subseteq \mathcal{T}_h$. For this work, we solely consider quadrilateral elements.
    \item[ii)] Embed $\Omega$ onto $\mathcal{T}_h$, and use some kind of identification, e.g. level sets, to locate elements outside $\Omega$ as $\mathcal{T}_h^{\text{out}}$, inside $\Omega$ as $\mathcal{T}_h^{\text{in}}$, or elements intersecting with the true boundary, $\Gamma$, as $\mathcal{T}_h^{\text{on}}$. With this, we have that $\mathcal{T}_h = \mathcal{T}_h^{\text{out}} \cup \mathcal{T}_h^{\text{in}} \cup \mathcal{T}_h^{\text{on}}$.
     \item[iii)] The unfitted computational surrogate domain is now chosen to consist of interior elements only, i.e., $\widetilde{\Omega}_h = \mathcal{T}_h^{\text{in}}$; however, including $\mathcal{T}_h^{\text{on}}$ is also a possibility.
\end{itemize}

\begin{figure}[h]
    \centering
    \includegraphics[scale = 0.9]{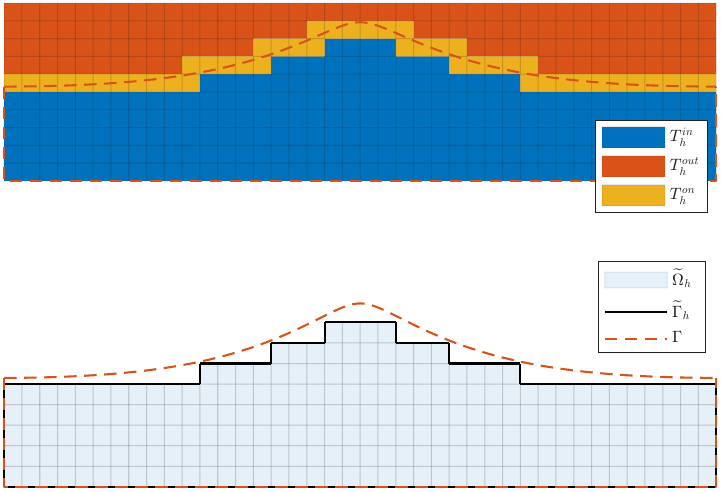}
    \caption{Illustration of the final unfitted computational domain (bottom plot) and the construction hereof (top plot). Depicted free surface profile of a highly nonlinear stream function wave ($kh = 1$, $\varepsilon/\varepsilon_{\max}=90\%$).}
    \label{fig:unfitted_domain_illustration}
\end{figure}

To support the understanding of the steps given above, consult Figure \ref{fig:unfitted_domain_illustration}. Here, the partitioning of $\mathcal{T}_h$ into $\mathcal{T}_h^{\text{out}}$, $\mathcal{T}_h^{\text{in}}$, and $\mathcal{T}_h^{\text{on}}$ is seen for a nonlinear free surface elevation profile, alongside, the final computational domain obtained using only $\widetilde{\Omega}_h = \mathcal{T}_h^{\text{in}}$ with the true and surrogate boundaries, $\Gamma$ and $\widetilde{\Gamma}_h$, respectively. 

Having introduced the unfitted computational domain, we have the following geometrical representation
\begin{equation}
     \widetilde{\Omega}_h  = \bigcup_{n = 1}^{N_{\text{elm}}} \mathcal{T}_h^n,
\end{equation}
with $\mathcal{T}_h^n$ being the $n$'th element from the initial background tesselation $\mathcal{T}_h$. Moreover, we denote by $\boldsymbol{x}=(x,z)$ a point on the true boundary $\Gamma$, and  $\widetilde{\boldsymbol{x}}=(\widetilde{x},\widetilde{z})$ a point on the surrogate boundary, $\widetilde{\Gamma}_h$. On $\Gamma$, we have the aforementioned outward-pointing normal, $\boldsymbol{n}$, and its associated tangent, $\boldsymbol{t} = (t_x,t_z) = (t_x(x,z,t),t_z(x,z,t)): \Gamma \mapsto \mathbb{R}^2$. Similarly on $\widetilde{\Gamma}_h$, the outward-pointing surrogate normal is defined as $\widetilde{\boldsymbol{n}} = (\widetilde{n}_x,\widetilde{n}_z) = (\widetilde{n}_x(\widetilde{x},\widetilde{z}),\widetilde{n}_z(\widetilde{x},\widetilde{z})): \widetilde{\Gamma}_h \mapsto \mathbb{R}^2$.

\subsection{The shifted boundary method}

% Introduce the shifted boundary method.
As mentioned in Section \ref{sec:introduction}, the shifted boundary methodology was first introduced by Main and Scovazzi in \cite{main2018shifted_part1,main2018shifted_part2} to solve boundary value problems with unfitted finite elements. The key idea is to solve the mathematical problem in question in the unfitted surrogate domain, $\widetilde{\Omega}_h$, instead of the true domain, $\Omega$. This is done by modifying the governing equations (typically through a weak Nitsche-type penalty approach), and \textit{shifting} the data of the boundary conditions accordingly between $\Gamma$ and $\widetilde{\Gamma}_h$. The shift is performed with two main ingredients: i) a unique mapping between true and surrogate boundary points and ii) Taylor series expansions of an appropriate order matching the intended approximation order, e.g., second order for the classical FEM. This has been shown (also for higher-order approximations) to preserve the expected optimal convergence order when applied consistently \cite{atallah2022high}. 

% The mapping:
As mentioned, a crucial ingredient in the shifted boundary approach is the mapping between the surrogate and true boundaries, $\widetilde{\Gamma}_h$ and $\Gamma$, defined as
\begin{equation}\label{eq:mapping}
     \mathcal{M}(\widetilde{\boldsymbol{x}}): \widetilde{\Gamma}_h \mapsto \Gamma, \quad \quad \quad \widetilde{\boldsymbol{x}} \mapsto \boldsymbol{x}.
\end{equation}
A lot of freedom is related to $\boldsymbol{x} = \mathcal{M}(\widetilde{\boldsymbol{x}})$. For this work, we use a purely vertical mapping for the free surface boundary, whereas for other boundaries (walls and bathymetry), we map along the true normal direction.

\subsection{Laplace problem in variational form}

For the sake of completeness, we consider a conformal boundary-fitted representation of $\Omega$ and $\Gamma$, given by $\Omega_h$ and $\Gamma_h$. In this setting, we want a variational formulation of the Laplace problem in \eqref{eq:Laplace_strong} in terms of an Aubin-type penalty method \cite{aubin1970approximation}. Here, we recast the strong formulation and weakly seek to find $\phi_h \in \mathcal{V}_h^P$ given
\begin{equation}
    a\langle \phi_h,v_h \rangle = b\langle v_h \rangle, \quad \forall v_h \in \mathcal{V}_h^P,
\end{equation}
where $a\langle \phi_h,v_h \rangle$ and $b\langle v_h \rangle$ denote the bilinear and linear forms, respectively, and $v_h$ is a test function from the following finite element space of piecewise continuous polynomials of degree $p$
\begin{equation}\label{eq:VP}
    \mathcal{V}^p_h = \{v_h \in C^0(\Omega); \forall n \in \{1,\ldots, N_{\text{elm}} \}, v_h |_{\mathcal{T}_h^n} \in \mathbb{P}^p(\mathcal{T}_h^n)\}.
\end{equation}
Furthermore, the bilinear and linear forms are given by
\begin{subequations}
\begin{align}
   a\langle \phi_h,v_h \rangle &= (\boldsymbol{\nabla} \phi_h, \boldsymbol{\nabla} v_h)_{\Omega_h} + \tau (\phi_h,v_h)_{\Gamma^{\eta}_h} - (\boldsymbol{\nabla} \phi_h \cdot \boldsymbol{n},v_h)_{\Gamma^{\eta}_h}, \\
   b\langle v_h \rangle &=  \tau (\phi_{\eta},v_h)_{\Gamma^{\eta}_h} + (q_i,v_h)_{\Gamma^{\text{i}}_h},
\end{align}
\end{subequations}
where $(a,b)_S = \int_S a \cdot b ~ dS$ is the classical $L^2$ inner product over some set $S$. Moreover, we have adopted implicit summation over $i \in \{\text{b},\text{w}\}$, with $q_i$ being the Neumann flux on the corresponding boundary. For this work, $q_i = 0$, but it is kept for the purpose of clarity. Note that periodic boundaries are imposed strongly by imposing continuity between the two domain ends. Lastly, $\tau$ is the penalty parameter that governs the stabilization term in the formulation.

Now, we make the change from a conformal to an unfitted mesh, i.e., $\Omega_h \rightarrow \widetilde{\Omega}_h$, and naively state the bilinear and linear form as
\begin{subequations}
\begin{align}
   a\langle \phi_h,v_h \rangle &= (\boldsymbol{\nabla} \phi_h, \boldsymbol{\nabla} v_h)_{\widetilde{\Omega}_h} + \tau (\phi_h,v_h)_{\widetilde{\Gamma}^{\eta}_h} - (\boldsymbol{\nabla} \phi_h \cdot \widetilde{\boldsymbol{n}},v_h)_{\widetilde{\Gamma}^{\eta}_h}, \\
   b\langle v_h \rangle &=  \tau (\widetilde{\phi}_{\eta},v_h)_{\widetilde{\Gamma}^{\eta}_h} + (\widetilde{q}_i,v_h)_{\widetilde{\Gamma}^{\text{i}}_h},
\end{align}
\end{subequations}
where the unknown Dirichlet and Neumann conditions are defined as
\begin{subequations}
\begin{align}
    \widetilde{\phi}_{\eta}    &= \phi_h(\widetilde{\boldsymbol{x}}),                                          & \text{on} \quad \widetilde{\Gamma}^{\eta}_h, \\
    \widetilde{q}_i    &= \boldsymbol{\nabla} \phi_h(\widetilde{\boldsymbol{x}}) \cdot \widetilde{\boldsymbol{n}},   &  \text{on} \quad   \widetilde{\Gamma}^{\text{i}}_h.
\end{align}
\end{subequations}
The next task is to determine the value of the boundary data, $\widetilde{\phi}_{\eta}$ and $\widetilde{q}_i$, on $\widetilde{\Gamma}_h$ using $\phi_{\eta}$ and $q_i$ from $\Gamma$. By simply equating those, the methodology is limited to first order; hence, a more elaborate exercise must be carried out. 

\subsection{Polynomial corrected shifted boundary approximation}

In the context of the original SBM, a second-order accurate truncated Taylor series is used to match the boundary data between the true and surrogate boundary, \cite{main2018shifted_part1,main2018shifted_part2}. Using only the mapping, $\mathcal{M}(\widetilde{\boldsymbol{x}})$, the SBM is able to enforce the correct boundary data, while preserving the right accuracy (second order) of the underlying finite element discretization. This approach has also been extended to high-order finite element spaces; see, e.g., \cite{atallah2022high}, where high-order Taylor series are truncated at the proper order. However, this poses an impractical feature, namely, the increased number of terms included in the series, as the order and especially the spatial dimension increase. Motivated by this, a simplified formulation was given in \cite{ciallella2022shifted}, exploiting a polynomial correction that is equivalent to a high-order Taylor series. This methodology was extended to the Poisson problem using very high-order basis functions in \cite{visbech2025spectral} and very recently generalized in \cite{ciallella2026generalized}. In the following, we use this methodology. Consult \cite{visbech2025spectral, ciallella2026generalized} for the derivation of the polynomial correction for Dirichlet and Neumann boundary conditions. Ultimately yielding
\begin{subequations}
\begin{align}
\widetilde{\phi}_{\eta}    &= \phi_{\eta} + \phi_h - \phi_h(\boldsymbol{x}), \quad &\text{on} \quad  \widetilde{\Gamma}_h^{\eta},  \\
\widetilde{q}_i    &= \widetilde{n}_{n} (q_i - \nabla \phi_h(\boldsymbol{x}) \cdot \boldsymbol{n} + \nabla \phi_h \cdot \boldsymbol{n})  + \widetilde{n}_{t} \nabla \phi_h \cdot \boldsymbol{t}, \quad &\text{on} \quad  \widetilde{\Gamma}_h^{i}, 
\end{align}
\end{subequations}
where the following decomposition of the surrogate normals has been used to reach the end result
\begin{equation}
    \widetilde{\boldsymbol{n}} = (\widetilde{\boldsymbol{n}} \cdot \boldsymbol{n}) \boldsymbol{n} + (\widetilde{\boldsymbol{n}} \cdot \boldsymbol{t}) \boldsymbol{t} = \widetilde{n}_{n}\boldsymbol{n} + \widetilde{n}_{t}\boldsymbol{t} ,
\end{equation}
and, most importantly, it can be observed that known boundary data, $\phi_{\eta}$ and $q_i$, from $\Gamma$ are imposed by evaluating solutions from the unfitted domain at the true boundary, $\phi_h(\boldsymbol{x})$. The final variation statement reads
\begin{subequations}\label{eq:Laplace_final_weak_form}
\begin{align}
   a\langle \phi_h,v_h \rangle = ~&(\boldsymbol{\nabla} \phi_h, \boldsymbol{\nabla} v_h)_{\widetilde{\Omega}_h} + \tau (\phi_h(\mathcal{M}(\widetilde{\boldsymbol{x}})),v_h)_{\widetilde{\Gamma}^{\eta}_h} - (\boldsymbol{\nabla} \phi_h \cdot \widetilde{\boldsymbol{n}},v_h)_{\widetilde{\Gamma}^{\eta}_h} \notag \\
   & + \widetilde{n}_{n} (\nabla (\phi_h(\mathcal{M}(\widetilde{\boldsymbol{x}}))-  \phi_h ) \cdot \boldsymbol{n},v_h)_{\widetilde{\Gamma}^{i}_h} - \widetilde{n}_{t} (\nabla \phi_h \cdot \boldsymbol{t},v_h)_{\widetilde{\Gamma}^{i}_h},  \\
   b\langle v_h \rangle =  ~&\tau (\phi_{\eta},v_h)_{\widetilde{\Gamma}^{\eta}_h} + \widetilde{n}_{n}(q_i,v_h)_{\widetilde{\Gamma}_h^{i}}
\end{align}
\end{subequations}
%\begin{subequations}\label{eq:Laplace_final_weak_form}
%\begin{align}
%   a\langle \phi_h,v_h \rangle = ~&(\boldsymbol{\nabla} \phi_h, \boldsymbol{\nabla} v_h)_{\widetilde{\Omega}_h} + \tau (\phi_h(\boldsymbol{x}),v_h)_{\widetilde{\Gamma}^{\eta}_h} - (\boldsymbol{\nabla} \phi_h \cdot \widetilde{\boldsymbol{n}},v_h)_{\widetilde{\Gamma}^{\eta}_h} \notag \\
   %& + \widetilde{n}_{n} (\nabla (\phi_h({\boldsymbol{x}})-  \phi_h ) \cdot \boldsymbol{n},v_h)_{\widetilde{\Gamma}^{i}_h} - \widetilde{n}_{t} (\nabla \phi_h \cdot \boldsymbol{t},v_h)_{\widetilde{\Gamma}^{i}_h},  \\
%   b\langle v_h \rangle =  ~&\tau (\phi_{\eta},v_h)_{\widetilde{\Gamma}^{\eta}_h} + \widetilde{n}_{n}(q_i,v_h)_{\widetilde{\Gamma}_h^{i}}
%\end{align}
%\end{subequations}
In the following we will simply use the notation $\boldsymbol{x} $
instead of $\mathcal{M}(\widetilde{\boldsymbol{x}})$, since on $\widetilde{\Gamma}_h$ we have by definition $\boldsymbol{x} = \mathcal{M}(\widetilde{\boldsymbol{x}})$.

\subsection{The spectral element method}\label{sec:SEM}

To represent any variable on the computational fluid domain, we employ a high-order spectral element method (SEM). The polynomial-based approach dates back to Patera's work in the 1980s \cite{patera1984spectral}. Since then, the method has established strong theoretical support \cite{karniadakis2005spectral} and has been used to model a wide range of problems \cite{xu2018spectral}. In the SEM, the global discrete solution, e.g., $\phi_h$, can be expressed as $N_{\text{elm}}$ local solutions, $\phi_h^n$, on each element as 
\begin{equation}\label{eq:poly_representation}
    \phi_h = \bigoplus_{n = 1}^{N_{\text{elm}}} \phi^n_h, \quad \text{where}\quad  \phi^n_h \approx \sum_{m=1}^{N_{\text{ep}}} \widehat{\phi}^n_{h,m} \mathcal{L}_{m}=\sum_{i=1}^{N_{\text{ep}}} \phi^n_{h,i} l_{i}.
\end{equation}
Here, each local solution is equivalently expressed using modal or nodal basis functions of up to degree $p$ from \eqref{eq:VP}. As we work on a Cartesian quadrilateral mesh, both bases are constructed as tensor products of 1D function spaces. Therefore, we have $N_{\text{ep}} = (1+p)^2$ quadrature points on each element distributed as Gauss-Lobatto-Legendre (GLL). The modal side is made up of its coefficients, $\widehat{\phi}^n_{h,m}$, and Legendre polynomials, $\mathcal{L}_{m}$. The nodal counterpart consists of the local solution values, $\phi^n_{h,i}$, and Lagrange polynomials, $l_i$. Following \cite{hesthaven2008nodal}, a nodal spectral element Galerkin discretization of the Laplace problem \eqref{eq:Laplace_final_weak_form} is obtained. This is done by defining elemental contributions and operations in terms of generalized Vandermonde matrices linking the nodal and modal expansions in \eqref{eq:poly_representation}.

By inserting our polynomial representation from \eqref{eq:poly_representation} into the final weak statement in \eqref{eq:Laplace_final_weak_form} and adopting a Galerkin formulation by choosing our test functions $v_h$, to equal our basis function, we construct a large linear system of equations, $\mathcal{A} \boldsymbol{\phi}_h = \boldsymbol{b}$. Here, $\mathcal{A} \in \mathbb{R}^{N^{\Omega}_{\text{dof}} \times N^{\Omega}_{\text{dof}}}$ is a sparse system matrix containing element and boundary contributions from the bilinear part of \eqref{eq:Laplace_final_weak_form}, $\boldsymbol{b} \in \mathbb{R}^{N^{\Omega}_{\text{dof}}}$ is a system vector containing boundary contributions from the linear part of \eqref{eq:Laplace_final_weak_form}, and $\boldsymbol{\phi}_h \in \mathbb{R}^{N^{\Omega}_{\text{dof}}}$ is the solution vector with $N_{\text{dof}}^{\Omega} = \text{DoF}(\widetilde{\Omega}_h)$. For this work, we solve the linear system directly in MATLAB (version R2025b) using a standard Gaussian elimination procedure.

\subsubsection{Spectral elements for fully nonlinear potential flow wave modeling}\label{sec:FNPF_SEM}

Using high-order SEM to model wave propagation and wave-structure interactions within the FNPF framework has been under active development for the past decades \cite{robertson1999free, engsigkarup2016stabilised, engsigkarup2019mixed,engsigkarup2019spectral,visbech2024solving,eskilsson2025co,visbech2025fnpf}, denoted FNPF-SEM in the following. As noted in \cite{kreiss1972comparison}, the key to computational efficiency for large-scale wave propagation, especially over long time durations, is high-order in accuracy. Modeling waves with an element-wise spectral expansion yields an accurate numerical solution with both $h$ and $p$ convergence strategies, and can be solved very efficiently; see, e.g., the $p$-multigrid approach in \cite{engsigkarup2021efficient}. Moreover, scalability can be achieved on multi-CPU systems as well \cite{eskilsson2025co,visbech2025fnpf}. Stabilization of the nonlinear free surface equations in \eqref{eq:free_surface_equations}, e.g., see Section \ref{sec:stabilization} in the following, is typically dealt with using a combination of over-integration and mild modal filtering \cite{engsigkarup2016stabilised}, and potentially re-meshing \cite{engsigkarup2019mixed}. Note that all present FNPF-SEM models have exact geometrical representation through either a $\sigma$-transformation or re-meshing of curvilinear elements. With this, stabilization is solely associated with i) the nonlinear terms from \eqref{eq:free_surface_equations} and ii) dealing with the intrinsic mesh-induced instability from asymmetric elements addressed in \cite{robertson1999free}. The solution for the latter is to use vertically aligned nodes near the free surface \cite{engsigkarup2016stabilised}. Nevertheless, almost all the proposed requirements from Section \ref{sec:introduction} are fulfilled for a reliable numerical model for FNPF wave modeling. The last untouched requirement is geometric flexibility, which, of course, is present in the FNPF-SEM methodology through boundary-fitted element tessellations. Yet if we want to include partially or fully submerged structures, which is impossible for \cite{engsigkarup2016stabilised} due to the $\sigma$-transform, it comes at a cost of re-meshing by adopting the non-transformed approach in \cite{engsigkarup2019mixed}. Here, the domain is re-meshed to account for the time-dependent curved free surface by vertically stretching a top layer of curvilinear quadrilaterals. This stretching effectively ruins $h$-convergence for highly nonlinear waves, as the vertical maximum element size is bounded by the difference between the minimum and maximum free-surface elevation. Arguably, for engineering purposes, errors at machine precision are a bit excessive, but from a theoretical standpoint, this feature is important to achieve. Motivated by this, we continue outlining our unfitted methodology. In Section \ref{sec:justification}, we justify using high-order approximations compared to lower-order ones, especially for long-time wave propagation.

\subsection{Fixed versus embedded discretizations}\label{sec:fixed_vs_embedded}

Modeling the FNPF wave problem with unfitted methodologies generally requires multiple objectives to be met. First and foremost, we need to capture the correct geometry of the problem (curved, complex, moving, deforming, etc.). When this is secured, we can focus on accuracy in both practical terms (e.g., engineering precision) and more theoretical terms. For the latter, and focusing solely on the free surface boundary, we need the embedded discretization as shown in Figure \ref{fig:unfitted_domain_illustration} to achieve a true fully $h$-convergent scheme. On the more practical side, the discretization strategy for the boundary-fitted FNPF-SEM models for pure wave propagation and interactions with bottom-mounted structures \cite{engsigkarup2016stabilised, engsigkarup2019mixed, engsigkarup2019spectral, eskilsson2025co, visbech2025fnpf} often uses polynomials of order $p\in\{3,\ldots,8\}$ on relatively coarse meshes in the vertical direction. Arguably, acting like a hybrid between a linear multi-domain approximation (classical FEM) and a pure spectral single-domain approximation (classical spectral method). Typically, applying $1 \leq N_z \leq 4$ elements in the vertical direction. Examples of this are given in the following:
In \cite{engsigkarup2016stabilised}, high-order harmonic generation over a submerged bar was performed using one single element, $N_z = 1$, in the depth of order $p=6$, with good visual comparison to experimental data \cite{beji1994numerical}.
In \cite{engsigkarup2019mixed}, a nonlinear soliton wall collision and reflection was modeled with excellent agreement compared to data from \cite{cooker1997reflection} using $N_z=1$ and $p=7$.
In \cite{engsigkarup2019spectral}, a nonlinear wave interaction with a FPSO-type structure using $N_z = 2$ and various orders $p \in \{3,4,5\}$. Ultimately, performing a comparable accuracy level as expensive CFD codes \cite{ransley2019blind}.
In \cite{eskilsson2025co}, a nonlinear shoaling test was performed using $N_z=1$ and $p=5$, showing good agreement with the experimental data.
In \cite{visbech2025fnpf}, modeling nonlinear wave-structure interactions with a vertical cylinder using $N_z = 4$ and $p = 5$, as well as wave interactions with a V-shaped breakwater. All showed very acceptable results when compared to other analytical/numerical/experimental data.

\begin{figure}[h]
    \centering
    \includegraphics{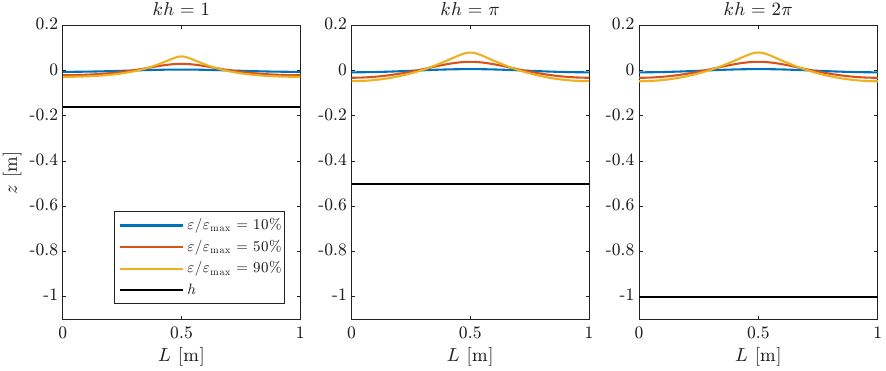}
    \caption{The relative depths, $h$, of three waves, $kh \in \{1, \pi, 2\pi\}$, with $\varepsilon/\varepsilon_{\max} \in \{10\%, 50\%, 90\%\}$.}
    \label{fig:three_waves}
\end{figure}

The pattern is clear: The practical requirement for modeling FNPF wave propagation with high-degree spectral elements is limited to about $1$ to $4$ elements in the vertical direction using a moderate to high polynomial order of the basis functions. The reason is related to the accuracy of sufficiently resolving the vertical dispersion relation, only requiring a few elements \cite{engsigkarup2009efficient, engsigkarup2013fast}. For the embedded discretization, depicted in Figure \ref{fig:unfitted_domain_illustration}, to be useful, practically speaking, the vertical length scale of the free surface, $l_{\eta} = \max(\eta) - \min(\eta)$, should at least be greater than the vertical length of the elements, $h_z$, near the free surface. In Figure \ref{fig:three_waves}, three waves of $kh \in \{1, \pi, 2\pi\}$ and a range of nonlinearity, $\varepsilon/\varepsilon_{\max} \in \{10\%, 50\%, 90\%\}$, are showcased with their associated depths, $h$. Clearly, deeper water (larger $kh$) yields a greater water column below the free surface, and higher nonlinearity increases the free-surface length scale, $l_{\eta}$. Now, checking how many elements can fit between $z = -h$ and $z = \min(\eta)$ for the three wave cases, before exceeding the limit of $h_z \leq 1.5l_{\eta}$, as an estimate of when embedded refinements start to be possible. The minimum number of elements in the vertical direction is reported in Table \ref{tab:three_waves_minimum_elements}, ultimately highlighting that for mild to moderate nonlinearity, the free surface embedding is not possible before a very fine mesh. For high nonlinearities, particularly in the shallow-water regime, fewer elements are needed before the limit is exceeded. The case using the fewest is for $kh = 1$, with $\varepsilon/\varepsilon_{\max} = 90 \%$, fitting $3$ elements, which is still within the vertical refinement level of the FNPF-SEM models reported in the literature.

\begin{table}[h]
\centering
\caption{Minimum number of elements in the vertical between $z = -h$ and $z = \min(\eta)$ not exceeding $h_z \leq 1.5l_s$.}\label{tab:three_waves_minimum_elements}
\begin{tabular}{r|rrr}
\hline
          & \multicolumn{3}{c}{$kh$} \\ \hline
$\varepsilon/\varepsilon_{\max}$ & $1$     & $\pi$    & $2\pi$    \\ \hline
$10 \%$         & 24    & 53    & 106    \\
$50 \%$       & 5     & 10    & 21     \\
$90 \%$        & 3     & 6     & 12     \\ \hline
\end{tabular}
\end{table}

Motivated by this, we introduce a \textit{fixed} unfitted discretization of $\widetilde{\Gamma}_h^{\eta}$ when the free surface surrogate boundary: i) is time-invariant, ii) forms a straight horizontal line, and iii) lies below $\min(\eta)$ at all times. Using the fixed discretization, we can still represent the moving, deforming, curved geometry of the free surface with reasonable accuracy in practical wave propagation problems, especially with higher polynomial orders.

%\textbf{Remark:} When the inclusion of fully or partially submerged structures is needed, additional work should be done to make this possible under a fixed domain discretization of the free surface. See our remark on this matter in Section \ref{sec:conclusion}.

\section{Numerical discretization: The free surface}\label{sec:numerical_dis_free_surface}

Next, we describe the numerical discretization of the free surface. To do this, we use a standard method-of-lines approach, in which the spatial discretization (employing a strong finite difference approximation) and the temporal discretization (using a standard explicit 4th-order, 4-stage Runge-Kutta scheme) are explained below, including considerations on gradient recovery and stabilization.

\subsection{Arbitrary-order finite difference method}\label{sec:finite_difference}

For the discretization of the nonlinear free surface equations in \eqref{eq:free_surface_equations}, we use classical finite difference operators as $\boldsymbol{\mathcal{D}}_x = \partial_x$. See the remark in Section \ \ref{sec:stabilization} on the choice of the discretization basis of the free surface. This approach has been employed to simulate FNPF wave problems \cite{li1997three, bingham2007accuracy, engsigkarup2009efficient}, where the free-surface conditions are approximated directly in strong form. We adopt the same nodal distribution as for spectral element discretizations, i.e., GLL points. Here, we assemble sparse finite difference operators of arbitrary order (at least matching $p$) on a non-uniform grid using Fornberg's recursive weighting algorithm. Typically, we use central stencils of size $r = \alpha + \beta + 1$, where $\alpha$ and $\beta$ denote the number of points used to the left and right of the central point, respectively. For central stencils, $\alpha = \beta$, and for finite domains, we employ off-centered stencils near the boundaries, thus $\alpha \neq \beta$. In semi-discrete form, we have
\begin{subequations}
\begin{align}
    \partial_t \boldsymbol{\eta} &= \boldsymbol{w}_{\eta} - \boldsymbol{\mathcal{D}}_x \boldsymbol{\eta} \boldsymbol{\mathcal{D}}_x \boldsymbol{\phi}_{\eta} +  \boldsymbol{w}_{\eta} \boldsymbol{\mathcal{D}}_x \boldsymbol{\eta} \boldsymbol{\mathcal{D}}_x \boldsymbol{\eta}, \\
    \partial_t \boldsymbol{\phi}_{\eta} &= - g \boldsymbol{\eta} - \frac{1}{2} \boldsymbol{\mathcal{D}}_x \boldsymbol{\phi}_{\eta} \boldsymbol{\mathcal{D}}_x \boldsymbol{\phi}_{\eta} +  \frac{1}{2} \boldsymbol{w}_{\eta}^2 (1 +  \boldsymbol{\mathcal{D}}_x \boldsymbol{\eta} \boldsymbol{\mathcal{D}}_x \boldsymbol{\eta}), 
\end{align}
\end{subequations}
where $\boldsymbol{\mathcal{D}}_x \in \mathbb{R}^{N^{\Gamma^{\eta}}_{\text{dof}} \times N^{\Gamma^{\eta}}_{\text{dof}}}$, $\{ \boldsymbol{\eta},\boldsymbol{\phi}_{\eta},\boldsymbol{w}_{\eta}\} \in \mathbb{R}^{N^{\Gamma^{\eta}}_{\text{dof}}}$, and $N_{\text{dof}}^{\Gamma^{\eta}} = \text{DoF}(\widetilde{\Gamma}_h^{\eta})$.

\subsection{Polynomial preserving gradient recovery}\label{sec:gradient_recovery}

% Motivation:
In the FNPF formulation, the vertical free surface velocity, $w_{\eta}$, serves as the dispersive link between the Laplace problem in 
\eqref{eq:Laplace_strong} and the two free surface conditions in 
\eqref{eq:free_surface_equations}. Here, $w_{\eta}$ is a crucial component for achieving an overall accurate numerical scheme and is therefore vital for the model's numerical stability. How we compute $w_{\eta}$ via gradient recovery is therefore of great importance. Gradient recovery, in the context of both low- and high-order finite elements, is a broad research area, with two main challenges being i) avoiding loss of accuracy when differentiating the primary variable (in this case, $\phi$) and ii) preventing discontinuities at element interfaces. Our focus in this work is on the polynomial-preserving recovery (PPR) introduced in \cite{naga2005polynomial}.

% Introduce local recovery:
Recall how $w_{\eta}$ is related to $\phi$, namely, $w_{\eta} = w |_{z=\eta} = \partial_z \phi |_{z=\eta}$. As a baseline, let us introduce local recovery: An efficient way to compute $w_{\eta}$ is to loop over the top layer of elements in the mesh and compute a local approximation of $w_h^n = D_z^n\phi_h^n$, where $D_z^n$ is a local differential operator for the $n$'th element acting on the elements associated solution values $\phi_h^n$. Then, evaluate $w_{\eta,h}^n = \mathcal{P}^n w_h^n$, where $\mathcal{P}^n$ is some local projection operator, e.g., for boundary-fitted domains, extracting the corresponding free surface values from the top of the element (i.e., a simple index operation), and for the unfitted domains considered in this work, extrapolating onto the true boundary. Although this approach is very efficient, it introduces the aforementioned two problems, i.e., loss of accuracy and discontinuities on element interfaces.

% Introduce PPR:
Now, let's introduce our PPR-inspired approach by assuming order $p$ of the finite element polynomial space in \eqref{eq:VP}. Then, considering each $i$ free surface grid point, one by one, given as $(x_i,z_i)$. For each point, we define a one-dimensional vertical least-squares polynomial of order $q=p+1$, denoted by $f(s)$ and expressed as
\begin{equation}\label{eq:least_squares_poly}
    f(s) = \sum_{j=0}^q c_j \psi_j(s), \quad \text{with} \quad s = z - z_i,
\end{equation}
and where $c_j$ is the $j$'th coefficient and $\psi_j(s)$ is the $j$'th basis function taken from a monomial basis, such that $\psi_j(s) = s^j$, i.e., $\psi_0(s) = 1$, $\psi_1(s) = s$, $\psi_2(s) = s^2$, etc. For the $q$'th order polynomial to be uniquely defined, we require $q+1$ sampling points, $\boldsymbol{s} = [s_1,\ldots,s_{q+1}]^T$, taking as the $q+1$ closest points to $z_i$ in the vertical direction including $z_i$. Also, we need sampling values, $\boldsymbol{\phi}$, of our velocity potential, $\phi_h$, at these points. We construct $f(s)$ such that the squared residual evaluated in $\boldsymbol{s}$ with respect to $\boldsymbol{\phi}$, is minimized, i.e., $\mathcal{J}(\boldsymbol{c}) =  |\boldsymbol{A}\boldsymbol{c} - \boldsymbol{\phi} |^2$, where
$\boldsymbol{c} = [c_0,\ldots,c_q]^T$ and the Vandermonde matrix as $\boldsymbol{A}_{jk} = \psi_{j}(s_k)$ for $j \in \{0,\ldots,q\}$ and $k \in \{1,\ldots,q+1\}$. Expanding $\mathcal{J}(\boldsymbol{c})$, taking the gradient with respect to $\boldsymbol{c}$, and setting it to zero, gives the following normal equations, which can be solved for $\boldsymbol{c}$
\begin{equation}\label{eq:normal_eq}
    \boldsymbol{A}^T \boldsymbol{A} \boldsymbol{c} = \boldsymbol{A}^T \boldsymbol{\phi}, \quad \quad \longleftrightarrow \quad \quad \boldsymbol{c} = (\boldsymbol{A}^T \boldsymbol{A})^{-1} \boldsymbol{A}^T \boldsymbol{\phi}.
\end{equation}
Now, writing out \eqref{eq:least_squares_poly} as
\begin{equation}
    f(s)  = c_0 + c_1 s + c_2 s ^2 + c_3 s ^3 + \cdots,
\end{equation}
and analytically differentiating with respect to $s$ gives
\begin{equation}
    g(s) = \frac{\partial f(s)}{\partial s} =  c_1 + 2c_2s + 3 c_3 s ^2 + \cdots,
\end{equation}
and finally, evaluating $g(s)$ at $z = z_i$, which means $s = 0$, results in $g(0) = c_1$. Using this, the gradient at the $i$'th free surface point $(x_i,z_i)$ is simply taken as the second element of the coefficient vector, $\boldsymbol{c}$, obtained from solving the small least-squares problem in 
\eqref{eq:normal_eq}. As we differentiate a $q=p+1$ polynomial, we expect the error to behave as $\mathcal{O}(h^{q=p+1})$, matching the rate of the primary variable, $\phi_h$. It can be observed that, compared to local recovery, the projection operations required on unfitted domains are naturally incorporated into the gradient recovery process. Additionally, the method can be considered meshless, as it operates only on a vertical cluster of points and is not limited to any specific numerical discretization method, such as FEM, SEM, FDM, HPC, etc. Therefore, it is a general approach that could enhance other FNPF models, which may experience suboptimal convergence due to gradient recovery. Finally, this method extends to 3D, as shown here, with no additional complexity beyond the need for more free-surface grid points.

% Relate to FNPF-SEM
In the FNPF-SEM context \cite{engsigkarup2016stabilised,engsigkarup2019mixed,visbech2025fnpf}, a global $L^2$ projection is typically used, which evidently loses one order of accuracy for the entire scheme, as with the local recovery. Improving this could be highly beneficial, as will be shown in Section \ref{sec:justification}, where we underline the importance of using gradient recovery with optimal convergence properties.

\subsection{Temporal  integration}\label{sec:time_integration}

We perform temporal integration of the free surface conditions using a standard explicit 4th-order, 4-stage Runge-Kutta (ERK4) scheme with a constant time step size, $\Delta t = T/200$, unless otherwise specified in the simulations. To accompany the ERK4 scheme, we estimate the time derivative of $\phi$ apparent in the force equation in \eqref{eq:force}, with a 4th-order-accurate finite difference approximation.

\subsubsection{Numerical stabilization}\label{sec:stabilization}

% Small review of existing stabilization approaches for FNPF wave solver:
When numerically integrating the FNPF free surface equations in \eqref{eq:free_surface_equations}, some form of numerical stabilization is often needed to prevent the simulation from becoming unstable, especially for highly nonlinear waves that are prone to aliasing-driven numerical instabilities over long integration times for such waves. The type of stabilization is related to the type of numerical approximation error introduced by the chosen discretization scheme. For explanation, generally, stability issues for FNPF wave problems can be divided into two categories related to the: i) numerical approximation and ii) geometrical representation. For the former, one can have a highly accurate numerical scheme. Yet, the simulation can still become unstable because aliasing errors accumulate in high-frequency components of the wave, eventually leading to undesirable behavior, e.g., saw-tooth-shaped instabilities \cite{longuet1976deformation}. For the latter, which is closely related to unfitted approaches, the instability is driven by errors arising from an inaccurate geometric representation. Adding too much stabilization will dampen the numerical solution; conversely, adding too little will cause the solution to become unstable and explode. Some examples of different stabilization approaches used within the field of FNPF wave simulation are given in the following:
% SEM
The aforementioned FNPF-SEM models, e.g., see \cite{engsigkarup2016stabilised,engsigkarup2019mixed,eskilsson2025co,visbech2025fnpf}, often apply a combination of i) over-integration of the nonlinear terms, ii) a mild spectral filter removing between 1-5$\%$ of the top modes in the interior of the free surface elements, and iii) some re-meshing to ensure the correct geometrical representation.
% FDM:
Wave models based on FDM, e.g., see \cite{bingham2007accuracy,engsigkarup2009efficient,xu2021finite} often use a very high-order Savitzky–Golay low-pass filter \cite{savitzky1964smoothing}. 
% HPC:
The original HPC work in \cite{shao2014harmonic} applied a 5-point low-pass filter at every 5-30 time steps, following the recommendations of \cite{longuet1976deformation}. Later HPC frameworks also used the Savitzky–Golay filter \cite{tong2024adaptive}.
% HOS:
The high-order spectral (HOS) model by \cite{dommermuth1987high} also used the 5-point low-pass filter in their original work. 

% Motivation and introducing hyperviosicty:
As we applied a finite-difference approximation to the free surface, we initially considered the Savitzky–Golay filter. However, this filter is restricted to equidistant grids and only works with central stencils. Thus, making it impossible to use for a finite wave tank with non-periodic boundaries, increased resolution near structures, and, as in our case, non-uniform mesh spacing from our GLL point distribution. Motivated by this and the work in \cite{shankar2018hyperviscosity, nielsen2020computational}, we consider adding hyperviscosity nodally to \eqref{eq:free_surface_equations} following \cite{xu2023high}. The general idea is to add dissipative terms to the free surface equations to target the high-frequency waves where the instability arises. Applications of modal hyperviscosity can, e.g., be seen in \cite{ducrozet2026hos}. Our modified free surface equations read
\begin{equation}
    \partial_t \eta = \mathcal{L}(w_{\eta}) + \mathcal{N}(\eta,\phi_{\eta},w_{\eta}) + \mathcal{H}^m_{\gamma}(\eta), \quad \text{and} \quad \partial_t \phi_{\eta} = \mathcal{L}(\eta) + \mathcal{N}(\eta,\phi_{\eta},w_{\eta}) + \mathcal{H}^m_{\gamma}(\phi_{\eta}), 
\end{equation}
where $\mathcal{L}()$ and $\mathcal{N}()$ are the classical linear and nonlinear terms and $\mathcal{H}()$ is the added hyperviscosity given as
\begin{equation}
    \mathcal{H}^m_{\gamma}(f) = (-1)^{m+1} \gamma \sqrt{\frac{g \pi}{\Delta x}}\left ( \frac{\Delta x}{\pi} \right )^{2m} \boldsymbol{\mathcal{D}}_{xx}^m f.
\end{equation}
Here $\Delta x$ is a measure of the grid spacing, $\gamma$ is a non-dimensional scale coefficient, $m$ is the hyperviscosity order, and $\boldsymbol{\mathcal{D}}_{xx}$ is the free surface Laplacian built using finite difference as described in Section \ref{sec:finite_difference}. Consult \cite{xu2023high} for various considerations and discussion on choosing the parameters, $m$ and $\gamma$. In Section \ref{sec:long_time_periodic}, we test different parameter configurations for long-time simulation of very nonlinear waves.

\textbf{Remark}: As mentioned at the beginning of this section, stability issues for FNPF wave problems are related to the approximation and geometrical representation of the problem. For boundary-fitted methodologies, a domain transformation (e.g., the $\sigma$-transform), re-meshing, etc., can ensure exact alignment between the true and computational domains. Evidently, this is not the case for unfitted methodologies. Even if the numerical and geometric results are convergent, small-scale errors may still arise as extrapolation errors or jumps across discontinuities from the staircase-like free surface approximation, e.g., see Figure \ref{fig:unfitted_domain_illustration}. With this, the stabilization techniques used must be capable of containing such errors and ultimately stably propagating the free-surface wave profile.
Based on experience (not shown), the authors have found that the strong FDM approximation of the free-surface equations is less sensitive to the geometric error introduced by the unfitted discretization than a weak SEM approximation. This ultimately yields a hybrid discretization, yet the overall scheme is, most importantly, of arbitrary order in accuracy regardless of the discretization. In 2D, this is trivial; however, in 3D, where the free surface may have complex geometric features in the $xy$-plane, special care is required to capture these features in an FDM setting. With this, we avoid introducing an additional weak statement of the free surface equations, and thereby performing a mass matrix inversion to compute the right-hand side of \eqref{eq:free_surface_equations}.
The authors stress that this observation, comparing FDM \cite{bingham2007accuracy,engsigkarup2009efficient} and SEM approximations \cite{engsigkarup2016stabilised} for wave propagation, is no general claim and not necessarily applicable to classical boundary-fitted models. It is solely an observation for the formulation in question for this work. 
\section{Results: Verification in a periodic wave flume}\label{sec:results_periodic}

For the first results section, we consider wave problems defined in a periodic wave flume. Here, we focus on verifying the model's convergence behavior and its ability (in terms of accuracy and stability) to simulate highly nonlinear waves over long time periods. For this, we give recommendations on how to set and choose the parameters for the added stabilization of hyperviscosity. Moreover, we attempt to justify the use of higher-order approximations over lower-order ones and underline the importance of proper gradient recovery at a given error level after long-time wave propagation. The periodic domain contains one wave of fixed length $L = 1~[\si{m}]$, irrespectively of wave number, $kh$, and level of nonlinearity, $\varepsilon/\varepsilon_{\max}$. Hence, the depth $h$ will vary with $kh$.

\subsection{Convergence under mesh-refinement}

As argued in Section \ref{sec:fixed_vs_embedded}, the embedded discretization of the free surface is mostly relevant for highly nonlinear waves. Therefore, we perform a classical convergence study under mesh-refinement for fixed polynomial orders, $p \in \{1,\ldots,5\}$, on three different waves, $kh \in \{1, \pi, 2\pi\}$. All of nonlinearity, $\varepsilon/\varepsilon_{\max} = 90 \%$. The background mesh is defined using $8 \leq N_x \leq 50$ elements in the horizontal direction, and $N_z$ depends on $kh$, such that the elements are uniform/square in size ($h_x / h_z \approx 1$). The resolution is increased successively, one element at a time, in the $x$-direction. As argued in, e.g., \cite{antonelli2026isogeometric}, the classical method of \textit{doubling} the resolution, i.e., $N_x \in \{1,2,4,\ldots\}$, when conducting convergence studies could create a biased mesh configuration (either in or unfavorable regarding the convergence of any numerical metric). The error is measured in the vertical free surface velocity as $\|w_{\eta}-w_{\eta,h}\|_{\infty}$, with the former denoting the true solution and the latter the numerical solution. 

\begin{figure}[H]
    \centering
    \includegraphics[scale=0.85]{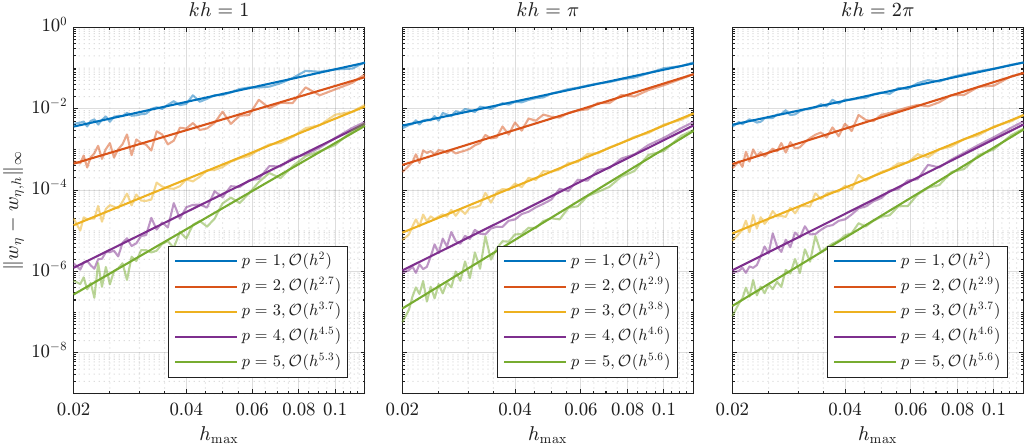}
    \caption{Convergence study under mesh-refinement for embedded discretizations $kh \in \{1, \pi, 2\pi\}$ with polynomial orders of $p \in \{1,\ldots,5\}$. The specified rate of convergence, $\mathcal{O}(h^m)$, is a linear fit to all data points.}
    \label{fig:h_convergence_embedded}
\end{figure}

The results can be seen in Figure \ref{fig:h_convergence_embedded}, where the error of each mesh is depicted with a linear fit of the convergence rate as $\mathcal{O}(h^m)$ (solid lines). We observe a highly mesh-dependent error level (transparent lines) that oscillates from one mesh to the next; however, it remains convergent. This behavior is well known in high-order unfitted methods. Slightly moving the true geometry on the same background mesh will, without a doubt, change the solution. This phenomenon can be seen, as mentioned, not only when shifting the boundary but also during successive mesh refinement \cite{antonelli2026isogeometric, ciallella2026generalized}. In general, we observe how all orders converge with at least $\mathcal{O}(h^p)$, most even closer to $\mathcal{O}(h^{p+1})$ (mostly true for $p \in \{1,2,3\}$). Also, the deep water waves, $kh \geq \pi$, have slightly better rates than the more shallow wave, $kh = 1$. The magnitude of errors is similar to results in the literature \cite{engsigkarup2016stabilised, engsigkarup2019mixed, visbech2025fnpf}, yet, as expected, slightly larger than boundary-fitted models. We emphasize the improved convergence rate of $\mathcal{O}(h^{p+1})$ for almost all orders of $p$ compared to other FNPF-SEM models. At finer resolutions, the error is more oscillatory than on coarse meshes (still convergent, though). This indicates that the error on the coarser meshes is primarily dominated by the numerical approximation of $\phi$ and $w_{\eta}$, rather than by the geometric representation of $\eta$. From this, we conclude that embedded discretizations are useful for highly nonlinear waves; however, if the accuracy in practical terms is expected, fixed discretizations can be employed as well.

\subsection{Justification for using high-order approximations and importance of optimal gradient recovery}\label{sec:justification}

In the following, we will attempt to justify the use of higher-order polynomials over lower-order ones and emphasize the importance of gradient recovery for optimal convergence properties. Considering the simplest possible case of a small-amplitude wave, i.e., $H/L \ll 1$, with $kh=2\pi$ (thus $h = L=1$ [\si{m}]) and analytical solutions from e.g., \cite{newman2018marine}. For this test, we introduce the relative error (in percent) in the free surface elevation, containing both numerical diffusion and dispersion effects, as 
\begin{equation*}
    e_{\text{rel}}(t) = \frac{\|\eta_h(t) - \eta(t) \|_{\infty}}{\|\eta(t) \|_{\infty}} \cdot 100 \%.
\end{equation*}
Next, we define the limit of "acceptable engineering accuracy" after $100$ wave periods as $10\%$, i.e., $e_{\text{rel}}(100T) \leq 10\%$. We want to locate the coarsest possible uniform mesh ($N_x = N_z = N$) for each polynomial order $p \in \{1,\ldots,6\}$, which satisfies this limit. Given this mesh, we can study the CPU times for the entire computation using two gradient recovery methods for $w_s$: PPR and local recovery, as discussed in Section \ref{sec:gradient_recovery}. For this test, we use $\Delta t = T/100$.

\begin{figure}[h]
    \centering
    \includegraphics[scale=0.85]{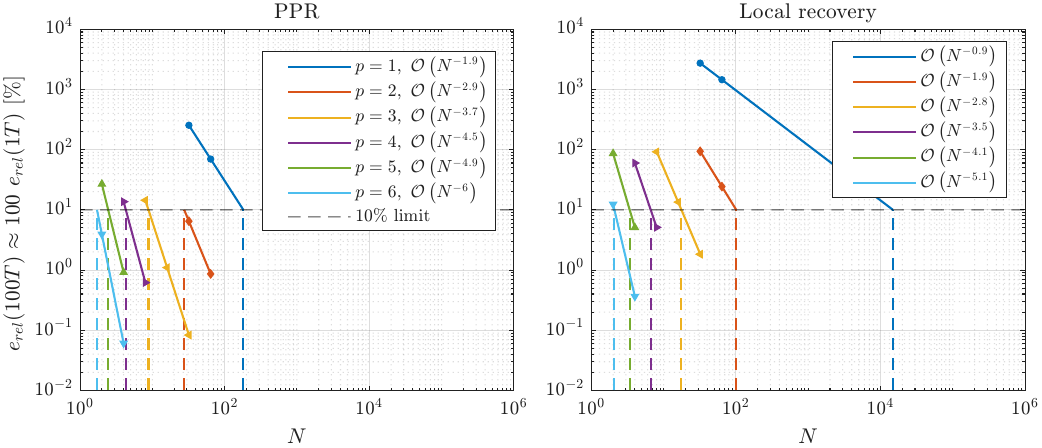}
    \caption{Estimation of the coarsest mesh configuration, $N$, to reach $e_{\text{rel}}(100T) \leq 10\%$ for $p \in \{1,\ldots,6\}$ and gradient recovery techniques: PPR (left) and local recovery (right).}
    \label{fig:high_order_justification_estimating}
\end{figure}

The coarsest possible resolution, given by $N$, was determined by running tests at various resolutions for only one wave period. Then, estimating and confirming the convergence rate, roughly $\mathcal{O}(h^{p+1})$ for PPR and $\mathcal{O}(h^{p})$ for local recovery, and predicting $N$ based on the acceptance limit by assuming that $e_{\text{rel}}(100T) \approx 100~e_{\text{rel}}(1T)$. This procedure is shown in Figure \ref{fig:high_order_justification_estimating} for PPR (left) and local recovery (right). Each $N$ is selected as the first integer above the non-integer, which arguably is a bit conservative. The coarsest mesh configurations, $N$, are listed in Table \ref{tab:high_order_justification}. From the figure, we generally see the expected difference in convergence order of one between PPR and local recovery. For $p \geq 4$, both methods are slightly below expectations, which may be attributed to the 4th-order-accurate ERK4 scheme and the fixed time step size. Despite this, the pattern is very clear when comparing the two recovery techniques. 

Now, running for each of those mesh configurations in Table \ref{tab:high_order_justification}, and measuring the CPU time for solely the time-integration part, those not including mesh-generation, building time-independent operators, post-processing, etc. As for computational hardware, we use an Intel Xeon E5-2660 v3 CPU at 2.60 GHz. Each computation is performed on a full node with 20 cores; however, it is run as a single-core computation with appropriate spin-up. From the table, it is clear that all runs were successful except for $p=1$ with local recovery, as this mesh has 220 million elements and is thus too demanding for a single-core computation. Next, we observe that all successful runs remain below the $10 \%$ limit as intended, where the one-wave-period estimate looks to have been more precise for finer meshes compared to coarser ones. Emphasizing how huge the difference is across the test: a 2 by 2 mesh with $p=6$ and PPR has a $3.03\%$ error after 100 wave periods and roughly 83 seconds of computations, whereas a 102 by 102 mesh with $p=2$ and local recovery has a $9.70 \%$ error after around $13,319$ seconds of computation. That equals a $1,593$ speed-up between the two tests.

\begin{figure}[h]
    \centering
    \includegraphics[scale=0.8]{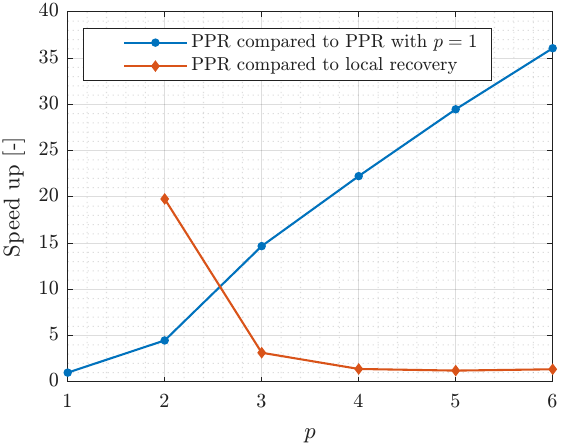}
    \caption{Speed up calculations of the results from Table \ref{tab:high_order_justification}. }
    \label{fig:high_order_justification_speed_up}
\end{figure}

At last, we consider relative speed-ups of the results in Table \ref{tab:high_order_justification} between i) PPR compared to PPR with $p=1$ and ii) PPR compared to local recovery with equal $p$. The results are shown in Figure \ref{fig:high_order_justification_speed_up}. Starting with i), we observe a speed-up of up to $\approx 36$ with $p=6$ compared to $p=1$. For ii) the speedup benefit looks to be mostly restricted to $p \leq 3$, which can be explained by the increased size of the least-squares problem in \eqref{eq:normal_eq} that increases with $p$ in both the number of basis functions and sampling points required. Finally, we stress that this small numerical study is not a general proof, yet still a reasonable indication and argument for i) why it is preferable to use higher-order approximations over lower ones, and ii) the importance of using optimal gradient recovery techniques. The presented speed-up times are comparable to those reported in \cite{engsigkarup2016stabilised}.

\begin{table}[h]
\centering \caption{Mesh input, $p$, $N$, $N_{\Omega_h}$, and DoF ($\Omega_h$), as a result of Figure \ref{fig:high_order_justification_estimating} and output, $e_{\text{rel}}(100T)$ and CPU time, for each run. "---" indicates an infeasible solution.}\label{tab:high_order_justification}
\begin{tabular}{c|rrrrrr}
\hline
\multicolumn{1}{l|}{Gradient recovery} & $p$ & $N$     & $N_{\Omega_h}$  & DoF ($\Omega_h$) & $e_{\text{rel}}(100T) ~[\%]$ & CPU time [\si{s}] \\ \hline
\multirow{6}{*}{PPR}                   & 1 & 182   & 33,124     & 33,306      & 9.05         & 3,017.07  \\
                                       & 2 & 28    & 784       & 3,192       & 8.61         & 673.54   \\
                                       & 3 & 9     & 81        & 756        & 9.11         & 205.55   \\
                                       & 4 & 5     & 25        & 420        & 4.70         & 135.63   \\
                                       & 5 & 3     & 9         & 240        & 3.70         & 102.38   \\
                                       & 6 & 2     & 4         & 156        & 3.03         & 83.64    \\ \hline
\multirow{6}{*}{Local}                 & 1 & 14,858 & 220,760,164 & 220,775,022  &  ---            & ---         \\
                                       & 2 & 102   & 10,404     & 41,820      & 9.70         & 13,319.30 \\
                                       & 3 & 18    & 324       & 2,970       & 9.62         & 647.47   \\
                                       & 4 & 7     & 49        & 812        & 8.26         & 189.25   \\
                                       & 5 & 4     & 16        & 420        & 5.03         & 124.73   \\
                                       & 6 & 3     & 9         & 342        & 1.60         & 113.87  \\ \hline
\end{tabular}
\end{table}

\subsection{Long time simulation with and without stabilization}\label{sec:long_time_periodic}

In the next part, we focus on long-time simulations of highly nonlinear waves on a fixed discretization. Recall from Section \ref{sec:stabilization}, we have two free parameters for our hyperviscosity stabilization, namely, the order $m$ and the scaling $\gamma$. In the following, we test a wide range of settings, i.e., $m = \{1,\ldots,5\}$ and $\gamma = \{0,0.1, 1, 10\}$, which are selected based on our personal experience with hyperviscosity stabilization, e.g., see \cite{xu2023high}. Note that $\gamma=0$ corresponds to running without any stabilization. We run the stabilization parameters for highly nonlinear $\varepsilon/\varepsilon_{\max} = 90\%$ waves of $kh  = \{1, \pi, 2\pi\}$ for $50T$ on meshes with $N_x = \{4,8,16\}$ horizontal elements and $N_z=\{1,2,4\}$ vertical elements of order $p = \{4,5,6\}$. This gives a total of $1,620$ combinations. Each run is categorized either as: i) unstable, if the numerical solution explodes, ii) too damped, if the numerical solution, in terms of $\eta$, becomes less than $50\%$ of the initial condition, and finally, iii) stable, otherwise. For all of the stable solutions, we compute a measure of relative diffusive and dispersive error (in percent) as
\begin{equation*}
    e_{\text{dif}}(t) =   \frac{|\underset{x}{\max}~ \eta_h(x,t)  - \underset{x}{\max}~\eta(x,t) | }{\underset{x}{\max} ~ \eta(x,t)}   \cdot 100 \%, \quad \text{and} \quad  e_{\text{dis}}(t) = \frac{| x^{\max}_h(t) - x^{\max}(t) |}{L} \cdot 100 \%,
\end{equation*}
where $x^{\max}_h(t)$ is the location of $\underset{x}{\max} ~ \eta_h(x,t)$ and similar for $x^{\max}(t)$. Note that $\eta_h$ is interpolated using Lagrangian interpolation to a very fine mesh of $\Delta x = 10^{-4}L$, whereon the two error measures are computed. 

In Table \ref{tab:best_runs_with_different_filters} in Appendix \ref{app:filter_runs}, the best 50 runs, based on minimization $e_{\text{mix}} = e_{\text{dif}}(50T) + e_{\text{dis}}(50T)$, are listed. Please note that from the 1,620 runs, 908 exploded, 225 were too damped, and 487 finished. Expecting all of the runs with $\gamma=0$ to explode, the success rate (in percent) is about $40 \%$, yet some waves will also be heavily dampened. From the table data, it is evident that different stabilization settings can be used, each working for various resolutions $(p,N_x,N_z)$ and waves. For example, out of the 50 best runs, the combination $(m,\gamma) = (3,1)$ occurs 13 times, $(m,\gamma) = (5,10)$ occurs 10 times, and $(m,\gamma) = (4,10)$ occurs 9 times. Moreover, it can be observed (as expected) that the runs with the best resolution give the best results, underlining the importance of having a well-resolved wave. Also, the $kh=1$ results appear to score best, because the water column below the free surface is better resolved at lower $kh$ than at higher $kh$. In the following, we use the parameters $m = 3$ and $\gamma = 1$ based on this test; however, other combinations are also valid. 

\begin{figure}[h]
    \centering
    \includegraphics[scale=0.9]{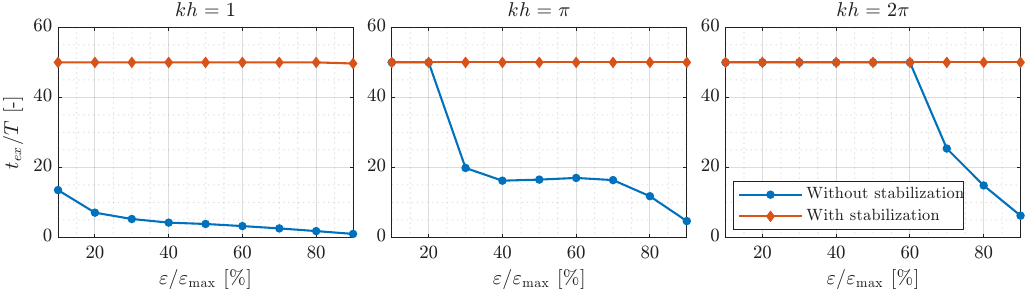}
    \caption{Time of explosion, $t_{ex}$, for three waves of $kh  = \{1, \pi, 2\pi\}$ at increasing nonlinearity, $\varepsilon/\varepsilon_{\max}$, with and without hyperviscosity stabilization using $(m,\gamma) = (3,1)$.}
    \label{fig:time_of_explosion}
\end{figure}

Now, applying these stabilization parameters, $(m,\gamma) = (3,1)$, to various waves, $kh  = \{1, \pi, 2\pi\}$, of different nonlinearities, $\varepsilon/\varepsilon_{\max} = \{10,20,\ldots,90\}$, all simulated on a $(N_x,N_z) = (8,2)$ mesh with $p = 6$ for a total time of $50T$. This resolution is chosen to enable a qualitative comparison with the analysis in \cite{engsigkarup2019mixed}, which uses a boundary-fitted SEM approach to solve the FNPF wave problem in a mixed-Eulerian-Lagrangian frame; however, they used only $N_z=1$. To make full use of our PPR technique to recover $w_{\eta}$, we need one additional element in the depth reaching $N_z=2$ in total. For completeness and as a reference, we also add a simulation without any stabilization, i.e., $\gamma = 0$. First, we consider the time of explosion, $t_{ex}$, if present, with and without stabilization as a function of nonlinearity. These results are reported in Figure \ref{fig:time_of_explosion}. From the figure, the simulations without stabilization explode in all three wave cases, whereas larger $kh$ yields a more stable solution for longer times than lower $kh$. Adding the stabilization of $(m,\gamma) = (3,1)$, makes the numerical solution stable until at least $50T$, expect for $kh=1$ and $\varepsilon/\varepsilon_{\max} = 90 \%$, which exploded after approximately $49.5T$, thus indicating that more resolution is needed combined with stabilization parameters targeting even higher modes, e.g., $(m,\gamma) = (4,10)$ or $(m,\gamma) = (5,10)$.

\begin{figure}[H]
    \centering
    \includegraphics[scale=0.85]{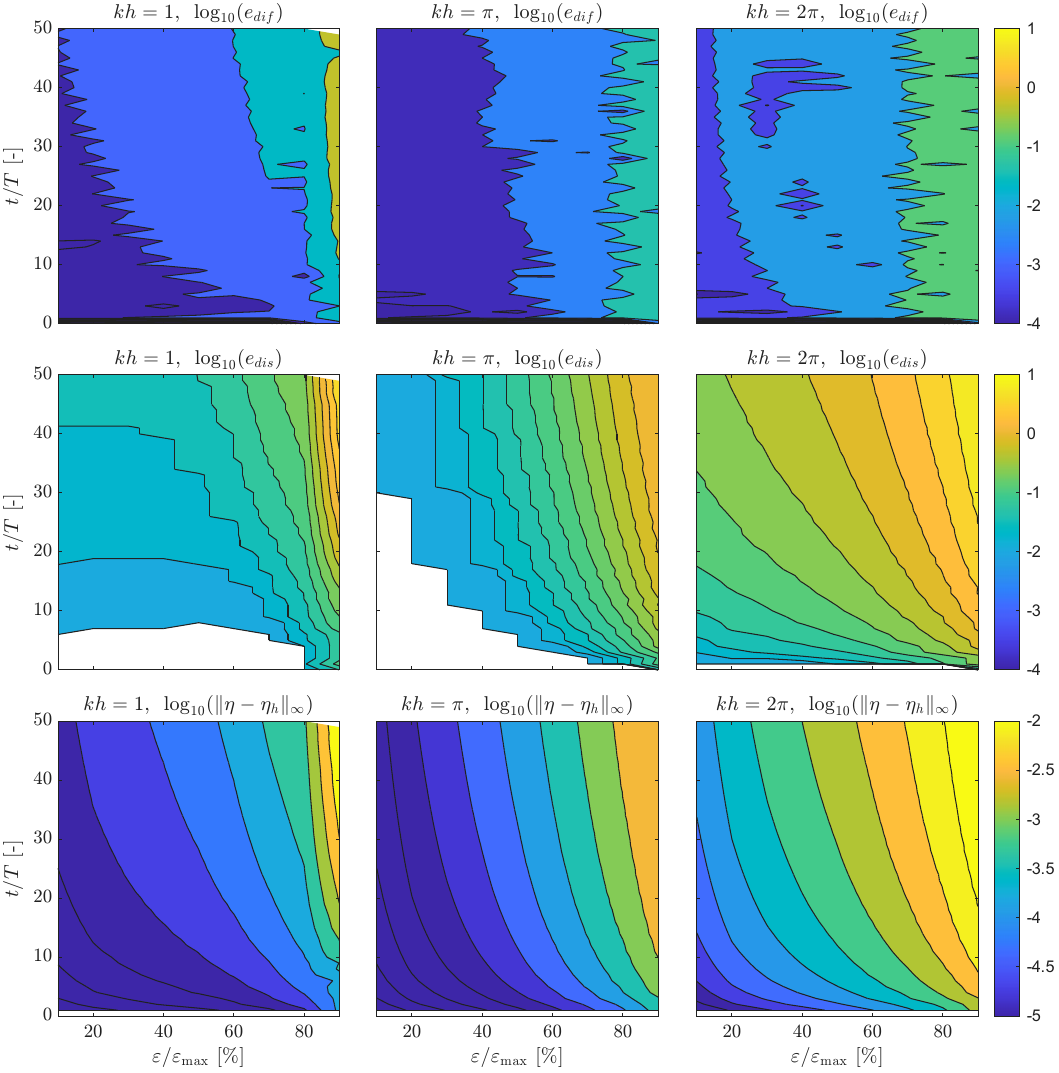}
    \caption{Time-dependent diffusive errors (top row), dispersive errors (middle row), and total error (bottom row) for different waves $kh  = \{1, \pi, 2\pi\}$ (columns) using a stabilization parameters of $(m,\gamma) = (3,1)$ on a mesh with $(N_x,N_z,p) = (8,2,6)$.}
    \label{fig:periodic_wave_contour}
\end{figure}

Next, in Figure \ref{fig:periodic_wave_contour}, the time-dependent diffusive, dispersive, and total errors are shown as a function of nonlinearity for the three different waves. We define the total error as $\|\eta - \eta_h \|_{\infty}$. As expected, all errors are larger with greater nonlinearity. In general, the diffusive errors are much lower (by 1-3 orders of magnitude) than the dispersive errors and appear to be nearly constant over time, underlining the low diffusiveness of high-order approximation methods. The white space in the dispersive errors is due to the fine mesh resolution used in the Lagrangian interpolation, with $\Delta x = 10^{-4}L$. In general, qualitative agreement with \cite{engsigkarup2019mixed} is evident.

\section{Results: Numerical experiments in a finite wave flume} \label{sec:results_finite}

For the next results, we consider wave problems in a finite-length wave flume that may be subject to bathymetric changes. This covers verification of wave generation and absorption zones, high-order harmonic generation over a submerged bar, and soliton interactions with a vertical wall and a semi-circular obstacle in the seabed. For all simulations, which vary in nonlinearity (from mild to highly nonlinear), we use the unfitted fixed-discretization approach; see Section \ref{sec:fixed_vs_embedded}.

\subsection{Wave generation and absorption on constant depth}

\begin{figure}[h]
    \centering
    \includegraphics[scale = 0.85]{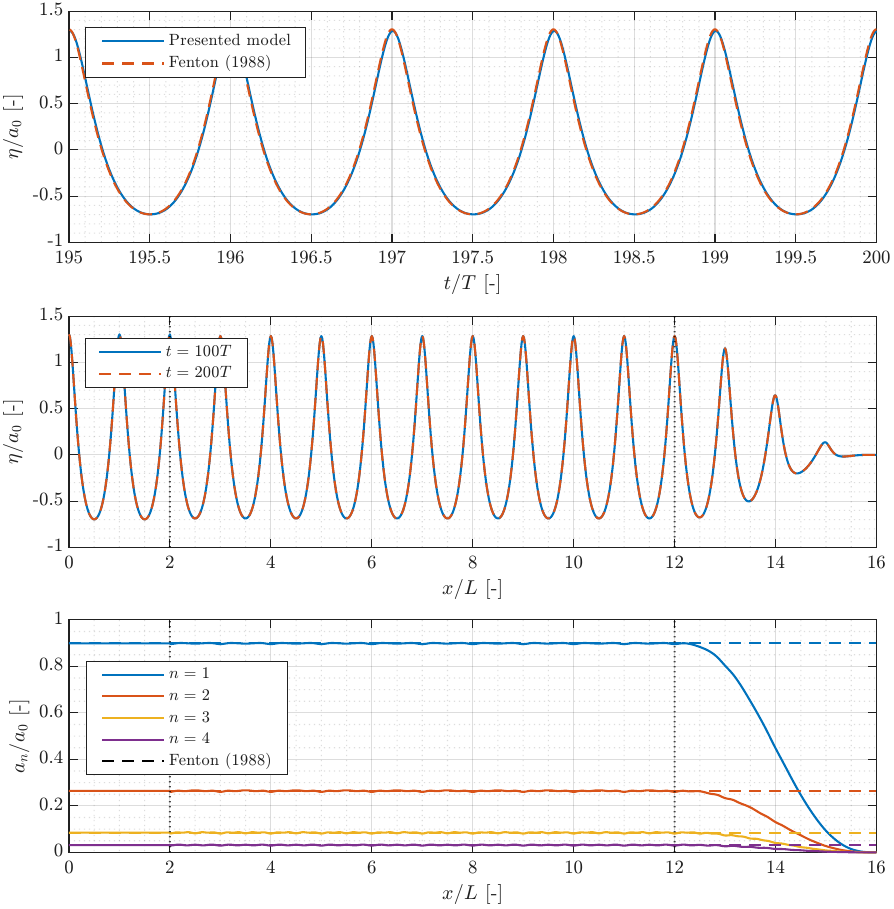}
    \caption{Simulation of a $kh = 1$ and $\varepsilon/\varepsilon_{\max} = 70 \%$ wave in a constant-depth wave tank with generation and absorption zones. Top: The last five wave periods evaluated at $x = 7L$ compared to \cite{fenton1988numerical}. Middle: Wave profile at 100 and 200 wave periods across the domain. Bottom: Harmonic analysis along the x-directions. For normalization, $a_0 = H/2 = 0.0351~[\si{m}]$.}
    \label{fig:finite_wave}
\end{figure}

First, we want to verify that our generation and absorption zones outlined in Section \ref{sec:ini_gen_abs} work as intended, especially for wave propagation over a very extensive time period. The computational domain is given by a generation zone ($0 \leq x \leq 2L$), followed by the target zone ($2L \leq x \leq 12L$), and an absorption zone at the end ($12L \leq x \leq 16L$), thus, the total length is $16L$. We consider a resolution of $N_x = 10$ per wavelength and $N_z = 4$ in the vertical direction, with basis functions of order $p = 6$. As for the wave input, we consider a wave with $kh = 1$ and $\varepsilon/\varepsilon_{\max} = 70 \%$ in nonlinearity. We employed $(m,\gamma)=(3,1)$ stabilization parameters. The simulation is performed for 200 wave periods. Temporal, spatial, and frequency results are shown in Figure \ref{fig:finite_wave}.

From the top plot, we see the wave free surface profile evaluated at the middle of the target domain, $x=7L$, for the last five wave periods of the simulation. Alongside this, we show the stream function solution from \cite{fenton1988numerical}. In general, there is reasonable visual agreement; however, a slight phase lag is evident. In the middle plot, we display the wave profiles along the domain at 100 and 200 wave periods, respectively. Overall, minimal to no differences are observed, indicating that all waves are being absorbed properly. Finally, a harmonic analysis is performed and shown in the bottom plot. This harmonic analysis is created by fitting (in a least-squares sense) a sum of trigonometric functions with the argument of $2\pi f_n t$, where $f_n = n/T$ for $n \in \{1,\ldots,4\}$. Additionally, $a_0$ is taken as $a_0 = H/2$. As expected, the agreement between the computed and expected results is acceptable.

\subsection{Harmonic generation over a submerged bar}\label{sec:bartest}

In the next experiment, we examine the classic and widely used validation test case of harmonic wave generation over a submerged bar \cite{beji1994numerical}. Here, waves are generated and then propagated toward the submerged bar, which transforms the wave by: i) steepening the wave due to nonlinear shoaling on the incline of the bar and ii) releasing bound harmonics as a decomposition of shorter "free" waves at the top of the bar. Ultimately, resulting in a completely transformed wave field behind the bar, overall testing the model's ability to model the nonlinear dispersion of waves.

The domain layout of the wave flume is as follows in $[\si{m}]$: Generation zone ($0 \leq x \leq 3$), target zone with the submerged bar ($3 \leq x \leq 25$), and absorption zone ($25 \leq x \leq 33$). The depth at the top of the bar is $h = 0.1~[\si{m}]$ and $h = 0.4~[\si{m}]$ away from the bar, with a 1:20 incline at the front and a 1:10 decline at the back. We used a nonlinear input wave with $kh = 0.675$, $T = 2.02~[\si{s}]$, and $\varepsilon/\varepsilon_{\max} \approx 7 \%$. For the computational inputs, we have employed a mesh with $(N_x,N_z) = (400,9)$ elements of order $p=4$ to accurately capture the bathymetric change. A comment on this is left at the end of Section \ref{sec:soliton_bump}. The simulation runs for $30$ period waves. For this test, no stabilization was needed.

\begin{figure}[h]
    \centering
    \includegraphics{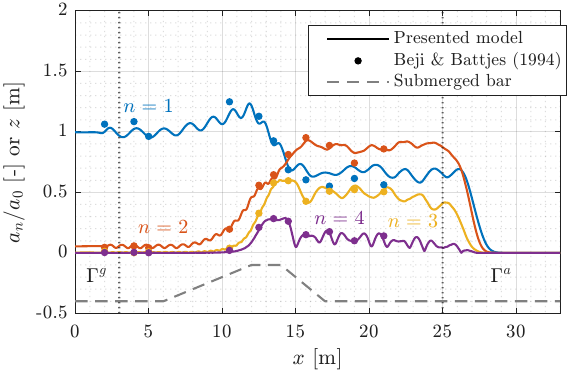}
    \caption{Harmonic analysis for the submerged bar test along the $x$-direction.}
    \label{fig:bar_test_harmonic}
\end{figure}

The time series of the free surface elevation, $\eta$, is given in Figure \ref{fig:bar_test_time_series} in Appendix \ref{app:bartest} at 10 different gauge locations governing the wave profiles before, on, and after the submerged bar. Clearly, the wave is steepened on the bar's incline and higher-order harmonics are generated and released at the incline, which is even more evident when considering the spectral components throughout the entire domain in Figure \ref{fig:bar_test_harmonic} alongside a visualization of the submerged bar. The analysis confirms the aforementioned observations and visually aligns well with the experimental results from \cite{beji1994numerical}.

\subsection{High-amplitude solitary wave interactions with a vertical wall}\label{sec:soliton_wall}

Next, we consider the nonlinear test of high-amplitude solitary waves reflecting off a vertical wall. This problem is well-studied, verified, and validated in the literature, with, e.g., experiments in \cite{maxworthy1976experiments} and numerical results reported in, e.g., \cite{cooker1997reflection, madsen2002new, engsigkarup2006nodal, klahn2021new} using different numerical approaches. We consider non-dimensionalized amplitudes as $0.075 \leq a/h \leq 0.6$, with $a$ being the initial soliton wave amplitude. For the test, we study four key metrics: i) maximum run-up height, $\eta_0$, occurring at time $t_0$, ii) attachment height, $\eta_a$, occurring before $t_0$ defined when the maximum instantaneous wave crest is located at the wall, iii) detachment height, $\eta_d$, occurring after $t_0$, defined when the crest leaves the wall, and iv) total (horizontal) force on the wall, $F$. In general, we note that $\eta_d < \eta_a < \eta_0$. When the soliton reaches the wall, a thin column of water will shoot up, and after the soliton leaves the wall, it will have reduced wave height and a small dispersive trail behind, making it a challenging nonlinear computational test. As $a/h$ increases and nonlinearity becomes more prominent, the expected run-up will no longer follow the linear elasticity relationship of $\eta_0/h \approx 2(a/h)$.

The domain and its computational layout are as follows: The domain is spanned by $ 0 \leq x \leq 40~[\si{m}]$ with a depth of $h = 1 ~[\si{m}]$. Initially, the center of the soliton is located at $x = 20~[\si{m}]$, and at $x = 40~[\si{m}]$, the vertical wall is located; thus, the soliton propagates from left to right. For the simulation, we use a constant time step of $\Delta t = 0.01~[\si{s}]$ and run until the wave has finished interacting with the wall. We have employed order $p=6$ basis functions on a mesh with $(N_x,N_z) = (100,2)$. For this test, no stabilization was used, as also noted in \cite{engsigkarup2006nodal,klahn2021new}.

\begin{figure}[h]
    \centering
    \includegraphics[scale = 0.9]{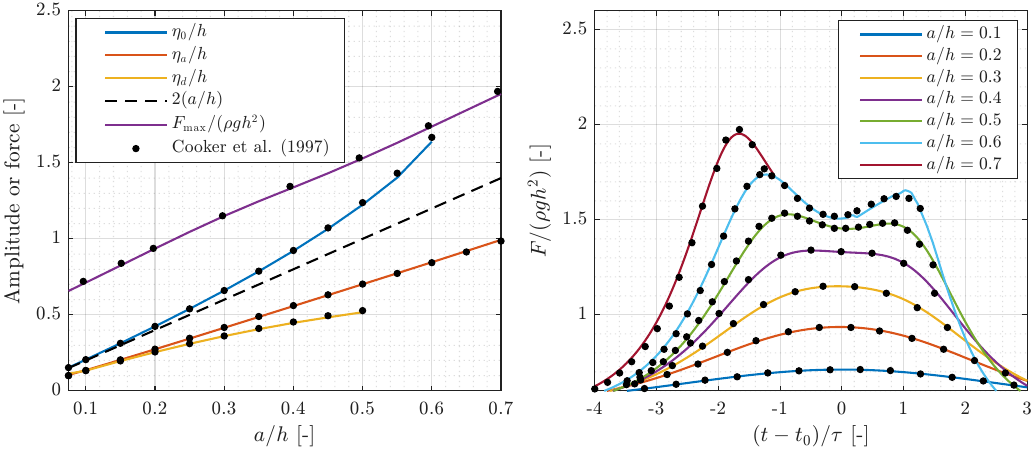}
    \caption{Solitary wave interactions with a vertical wall of different initial amplitudes, $a/h$. Left: maximum run-up height, $\eta_0$, attachment height, $\eta_a$, detachment height, $\eta_d$, and maximum force, $F_{\max}$. Right: total horizontal force signals, $F$, for different $a/h$. All results compared with numerical solutions from \cite{cooker1997reflection}.}
    \label{fig:soliton_wall}
\end{figure}

The results are seen in Figure \ref{fig:soliton_wall} with comparison to digitized values from the numerical results found in \cite{cooker1997reflection}. From the figure (left), we observe the maximum run-up height, $\eta_0$, attachment height, $\eta_a$, detachment height, $\eta_d$, and maximum horizontal force, $F_{\max}$, all showing good agreement. Here, we note that the inelastic interaction becomes visually apparent for $a/h > 0.2~[\text{-}]$. For the figure (right), we have the total horizontal force signals, also showing reasonable visual agreement with \cite{cooker1997reflection}. For this computation, the force is composed of three contributions: two from the vertical elements in the computational domain, and one from an artificial spectral element spanning the top of the fixed discretization and the free-surface elevation. Pressure values are projected via Lagrangian extrapolation onto the artificial element and integrated numerically as with the two other contributions. Upon closer inspection of the results for the second most nonlinear case, $a/h=0.6~[\text{-}]$, a small "kink" in the force signal can be observed just after the point of maximum run-up. These small discrepancies might be explained by employing a fairly extreme projection, i.e., extrapolating a value approximately $1.7/0.5 = 3.4$ element lengths away.
Regarding the most nonlinear case of $a/h=0.7~[\text{-}]$, we were only able to predict the attachment height, initial part of the force profile, and the maximum force. For this case, the nonlinearity is so great that the topology of the free surface elevation profile changes, i.e., non-single values for $\eta$, thus the simulation breaks down. The force signal in Figure \ref{fig:soliton_wall} (right) for $a/h=0.7~[\text{-}]$ is manually shifted as $t_0$ was unknown for this simulation. The authors note that stabilization could slightly improve the results for $a/h > 0.6$; however, the instability here is primarily expected to arise from the non-uniqueness of $\eta$ near the wall under extreme wave-wall interaction.

\subsection{Solitary wave propagation over a semi-circular bathymetry bump}\label{sec:soliton_bump}

As a final numerical study, we consider an experiment in which a mildly nonlinear solitary wave is propagated over a semi-circular bump in the bathymetry. The domain spans $0 \leq x \leq 45~[\si{m}]$ with a constant depth (except on the bump) of $h = 1~[\si{m}]$. The bump of radius $R = 0.3 ~[\si{m}]$ is located at $(x,z) = (30,-1) ~[\si{m}]$. The soliton is of height $a/h=0.2~[\text{-}]$ and initialized at $x = 15~[\si{m}]$. 

For the computation, a time step of $\Delta t = 0.01 ~[\si{s}]$ is used, and the simulation is run until the soliton has finished interacting with the bump. As with the experiment in Section \ref{sec:soliton_wall}, no stabilization was needed. The domain is uniformly meshed using $(N_x,N_z) = (450,11)$ elements with $p=3$. A zoom in at the unfitted mesh around the bump is shown in Figure \ref{fig:unfitted_semi_circular_bump} in Appendix \ref{app:mesh_semi_circular_bump}, with $14$ surrogate boundary elements approximating the curved semi-circle.

\begin{figure}[h]
    \centering
    \includegraphics[scale=1]{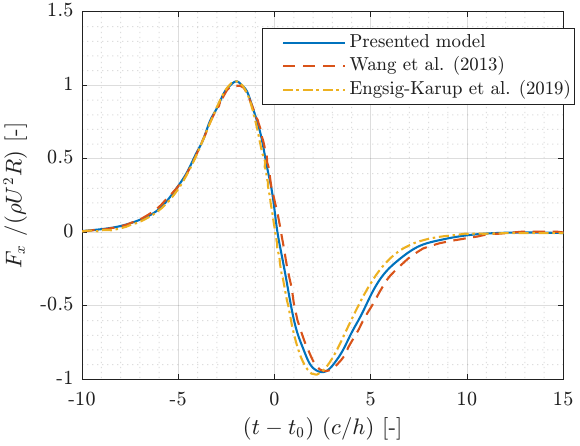}
    \caption{Total horizontal dynamic force on the semi-circular bump due to a propagating solitary wave.}
    \label{fig:soliton_semi_circular_bump}
\end{figure}

The results are shown in Figure \ref{fig:soliton_semi_circular_bump} together with the digitized numerical FNPF results of \cite{wang2013study,engsigkarup2019mixed} in terms of the total horizontal dynamic force, $F_x$, thus dropping $gz$ from \eqref{eq:force}. The force is computed by projecting the pressure onto the true boundary via Lagrangian interpolation and evaluating the integral numerically using artificial boundary-fitted curves. For the non-dimensionalization of the results, we have used the initial soliton wave speed, $U = 3.4276 ~[\si{m/s}]$, from \cite{dutykh2014efficient}, and the solitary wave phase celerity, $c = \sqrt{1+a/h} = 1.0954 ~[\si{m/s}]$. As defined in \cite{engsigkarup2019mixed}, $t_0$ is when the soliton is completely on top of the bump, experiencing zero horizontal force. In general, the computed results look to agree qualitatively well.

As may be obvious, the meshing of this particular problem is quite excessive with 4,928 elements of order $p=3$. Comparing with the boundary-fitted model in \cite{engsigkarup2019mixed}, where 164 elements of order $p=6$ were used. Similarly, this can be noted for the mesh used in Section \ref{app:bartest} using 3,100 elements
of order $p=4$, compared to 103 elements of order $p=6$ in \cite{engsigkarup2016stabilised}. When applying a uniform, structured background mesh, the far-field resolution (often requiring only one or two elements in the depth) depends on the length scale of the geometrical features (the submerged bar and the semi-circular bump) in the domain. That is especially true for the vertical direction. In Section \ref{sec:conclusion}, we discuss potential solutions to this challenge to further improve the model's efficiency.
\section{Conclusion}\label{sec:conclusion}

We have introduced a novel unfitted high-order computational framework for simulating highly nonlinear water waves that may be affected by bathymetric changes. The framework's key feature is the unfitted method for discretizing and representing the fluid domain, based on a polynomial-corrected shifted-boundary approximation and spectral elements. Additionally, a polynomial-preserving recovery technique was employed to achieve optimal convergence when computing the free-surface vertical velocity, thereby further improving the accuracy of the numerical scheme. We described the continuous mathematical formulation, followed by the discretization of the fluid domain and the free surface. For the former, the unfitted polynomial-corrected shifted-boundary method on spectrally accurate elements was presented, while for the latter, emphasis was placed on the polynomial-preserving gradient-recovery technique and wave stabilization. Subsequently, a series of numerical verification and validation results was presented, ultimately demonstrating the capabilities and applicability of this new unfitted wave model for simulating highly nonlinear stream function waves and high-amplitude solitons over extended periods with high numerical accuracy. In this section, significant efficiency improvements were highlighted when using higher-order discretizations compared with lower-order ones, especially when combined with optimal gradient recovery. 

% Future perspectives:
The proposed model (ongoing work) aims to be extended to include partially and/or fully submerged structures, whether fixed, forced, or eventually freely floating/oscillating. This emphasizes the need for geometric flexibility since multiple length scales will be present in the problem, as highlighted in Section 
\ref{sec:soliton_bump}. To address this, we must efficiently capture these scales with optimal resolution throughout the entire fluid domain. The proposed unfitted methodology could be combined with a non-conformal element discretization (i.e., two elements share the same element edge), i.e., a non-structured mesh approach to regular Cartesian grids, ultimately providing the geometric adaptivity necessary.

\section*{Acknowledgments}
The research was carried out at the Technical University of Denmark (DTU) in the Department of Applied Mathematics and Computer Science (Compute), with computational resources provided by the DTU Computing Center (DCC). Parts of the research were carried out during JV's visit to the CARDAMOM team at the  INRIA center at the University of Bordeaux. MR is a member of the CARDAMOM team at the INRIA center at the University of Bordeaux.

\section*{Declarations}
The authors declared that they had no potential conflicts of interest with respect to the research, authorship, and/or publication of this article. Generative AI was used solely to improve the grammar, spelling, and readability of this manuscript. The authors reviewed and edited all AI-assisted suggestions and take full responsibility for the final content.

\section*{Funding}
This work contributes to the activities of the PhD project of JV: ‘‘New Advanced Simulation Techniques for Wave Energy Converts’’, fully funded by DTU Compute. 

\section*{CRediT author contribution statement}
All: Conceptualization, Methodology, Writing - Original Draft, Writing - Review \& Editing, Visualization
Additional individual contributions:
\textbf{JV}: Software, Validation, Formal analysis, Investigation.
\textbf{APEK}, \textbf{HBB}, and \textbf{MR}: Supervision.

\bibliographystyle{SageH}
% \bibliographystyle{abbrvnat}
% \bibliographystyle{unsrtnat}

% When working on the draft use the following:
\bibliography{references.bib}  

@article{antonelli2026isogeometric,
  title={Isogeometric multipatch coupling with arbitrary refinement and parametrization using the Gap--Shifted Boundary Method},
  author={Antonelli, Nicol{\`o} and Gorgi, Andrea and Zorrilla, Rub{\'e}n and Rossi, Riccardo},
  journal={Computer Methods in Applied Mechanics and Engineering},
  volume={456},
  pages={118913},
  year={2026},
  publisher={Elsevier}
}

@article{atallah2022high,
title = {The high-order Shifted Boundary Method and its analysis},
journal = {Computer Methods in Applied Mechanics and Engineering},
volume = {394},
pages = {114885},
year = {2022},
author = {Nabil M. Atallah and Claudio Canuto and Guglielmo Scovazzi},
}

@article{atallah2025high,
  title={A high-order Shifted Interface Method for {L}agrangian shock hydrodynamics},
  author={Atallah, Nabil M and Mittal, Ketan and Scovazzi, Guglielmo and Tomov, Vladimir Z},
  journal={Journal of Computational Physics},
  volume={523},
  pages={113637},
  year={2025},
  publisher={Elsevier}
}

@article{aubin1970approximation,
	author = {Jean-Pierre Aubin},
	journal = {Journal of Mathematical Analysis and Applications},
	number = {3},
	pages = {510-521},
	title = {Approximation des probl{\`e}mes aux limites non homog{\`e}nes pour des op{\'e}rateurs non lin{\'e}aires},
	volume = {30},
	year = {1970}}

@inproceedings{amini2019nonlinear,
  title={A nonlinear potential-flow model for wave-structure interaction using high-order finite differences on overlapping grids},
  author={Amini-Afshar, Mostafa and Bingham, Harry B and Henshaw, William D and Read, Robert},
  booktitle={Proceedings: The 34th International Workshop on Water Waves and Floating Bodies, Newcastle, Australia},
  year={2019}
}

@article{amini2017solving,
    title = {{Solving the linearized forward-speed radiation problem using a high-order finite difference method on overlapping grids}},
    year = {2017},
    journal = {Applied Ocean Research},
    author = {Amini-Afshar, Mostafa and Bingham, Harry B},
    pages = {220--244},
    volume = {69},
}

@article{amini2018pseudo,
    title = {{Pseudo-impulsive solutions of the forward-speed diffraction problem using a high-order finite-difference method}},
    year = {2018},
    journal = {Applied Ocean Research},
    author = {Amini-Afshar, Mostafa and Bingham, Harry B},
    pages = {197--219},
    volume = {80},
}

@article{berger2024new,
  title={A new provably stable weighted state redistribution algorithm},
  author={Berger, Marsha and Giuliani, Andrew},
  journal={SIAM Journal on Scientific Computing},
  volume={46},
  number={5},
  pages={A2848--A2873},
  year={2024},
  publisher={SIAM}
}

@article{birke2023dod,
author = {Birke, Gunnar and Engwer, Christian and May, Sandra and Streitbürger, Florian},
title = {{DoD} Stabilization of linear hyperbolic {PDE}s on general cut-cell meshes},
journal = {Proceedings in Applied Mathematics and Mechanics},
volume = {23},
number = {1},
pages = {e202200198},
year = {2023}
}

@article{badia2018aggregated,
  title={The aggregated unfitted finite element method for elliptic problems},
  author={Badia, Santiago and Verdugo, Francesc and Mart{\'\i}n, Alberto F},
  journal={Computer Methods in Applied Mechanics and Engineering},
  volume={336},
  pages={533--553},
  year={2018},
  publisher={Elsevier}
}

@article{beji1994numerical,
title = {Numerical simulation of nonlinear wave propagation over a bar},
journal = {Coastal Engineering},
volume = {23},
number = {1},
pages = {1-16},
year = {1994},
author = {S. Beji and J.A. Battjes}
}

@article{burman2015cutfem,
  title={Cut{FEM}: discretizing geometry and partial differential equations},
  author={Burman, Erik and Claus, Susanne and Hansbo, Peter and Larson, Mats G and Massing, Andr{\'e}},
  journal={International Journal for Numerical Methods in Engineering},
  volume={104},
  number={7},
  pages={472--501},
  year={2015},
  publisher={Wiley Online Library}
}

@article{bingham2007accuracy,
    title = {{On the accuracy of finite-difference solutions for nonlinear water waves}},
    year = {2007},
    journal = {Journal of Engineering Mathematics},
    author = {Bingham, Harry B and Zhang, Haiwen},
    number = {1},
    pages = {211--228},
    volume = {58},
}

@article{cutforth2021efficient,
title = {An efficient moment-of-fluid interface tracking method},
journal = {Computers \& Fluids},
volume = {224},
pages = {104964},
year = {2021},
author = {Murray Cutforth and Philip T. Barton and Nikos Nikiforakis}
}

@article{cooker1997reflection,
  title={Reflection of a high-amplitude solitary wave at a vertical wall},
  author={Cooker, Mark J and Weidman, Patrick D and Bale, Derek S},
  journal={Journal of Fluid Mechanics},
  volume={342},
  pages={141--158},
  year={1997},
  publisher={Cambridge University Press}
}

@article{ciallella2026generalized,
      title={Generalized high-order minimization-based polynomial corrections on unfitted spectral elements for the Poisson problem}, 
      author={Mirco Ciallella and Jens Visbech},
      year={2026},
    journal = {arXiv:123.456}
}

@article{ciallella2022shifted,
title = {Shifted boundary polynomial corrections for compressible flows: high order on curved domains using linear meshes},
journal = {Applied Mathematics and Computation},
volume = {441},
pages = {127698},
year = {2023},
author = {Mirco Ciallella and Elena Gaburro and Marco Lorini and Mario Ricchiuto}
}

@article{chesshire1990composite,
  title={Composite overlapping meshes for the solution of partial differential equations},
  author={Chesshire, G and Henshaw, William D},
  journal={Journal of Computational Physics},
  volume={90},
  number={1},
  pages={1--64},
  year={1990},
  publisher={Elsevier}
}

@article{colomes2021weighted,
title = {A weighted Shifted Boundary Method for free surface flow problems},
journal = {Journal of Computational Physics},
volume = {424},
pages = {109837},
year = {2021},
author = {Oriol Colomés and Alex Main and Léo Nouveau and Guglielmo Scovazzi},
}

@article{dutykh2014efficient,
  title={Efficient computation of steady solitary gravity waves},
  author={Dutykh, Denys and Clamond, Didier},
  journal={Wave Motion},
  volume={51},
  number={1},
  pages={86--99},
  year={2014},
  publisher={Elsevier}
}

@article{davidson2020efficient,
  title={Efficient nonlinear hydrodynamic models for wave energy converter design—A scoping study},
  author={Davidson, Josh and Costello, Ronan},
  journal={Journal of Marine Science and Engineering},
  volume={8},
  number={1},
  pages={35},
  year={2020},
  publisher={MDPI}
}

@book{dean1991water,
  title={Water wave mechanics for engineers and scientists},
  author={Dean, Robert G and Dalrymple, Robert A},
  year={1991},
  publisher={World Scientific Publishing Company}
}

@article{duprez2023new,
  title={A new $\phi$-FEM approach for problems with natural boundary conditions},
  author={Duprez, Michel and Lleras, Vanessa and Lozinski, Alexei},
  journal={Numerical Methods for Partial Differential Equations},
  volume={39},
  number={1},
  pages={281--303},
  year={2023},
  publisher={Wiley Online Library}
}

@article{dommermuth1987high,
  title={A high-order spectral method for the study of nonlinear gravity waves},
  author={Dommermuth, Douglas G and Yue, Dick KP},
  journal={Journal of Fluid Mechanics},
  volume={184},
  pages={267--288},
  year={1987},
  publisher={Cambridge University Press}
}

@article{ducrozet2026hos,
  title={{HOS-NWT}: an open-source numerical wave tank based on the High-Order Spectral method},
  author={Ducrozet, Guillaume and Charlou, Moran and Bonnefoy, F{\'e}licien},
  journal={Computer Physics Communications},
  pages={110168},
  year={2026},
  publisher={Elsevier}
}

@article{eskilsson2025co,
  title={A co-simulation approach to modelling fully nonlinear water waves using the spectral/hp element method},
  author={Eskilsson, Claes and Abedi, Hamidreza and Engsig-Karup, Allan P},
  journal={Ocean Engineering},
  volume={340},
  pages={122385},
  year={2025},
  publisher={Elsevier}
}

@article{evans1987report,
  title={Report on the first international workshop on water waves and floating bodies},
  author={Evans, DV and Newman, JN},
  journal={Journal of Fluid Mechanics},
  volume={174},
  pages={521--528},
  year={1987},
  publisher={Cambridge University Press}
}

@article{engsigkarup2006nodal,
  title={Nodal {DG-FEM} solution of high-order {B}oussinesq-type equations},
  author={Engsig-Karup, Allan P and Hesthaven, Jan S and Bingham, Harry B and Madsen, Per A},
  journal={Journal of Engineering Mathematics},
  volume={56},
  number={3},
  pages={351--370},
  year={2006},
  publisher={Springer}
}

@article{engsigkarup2019spectral,
title = "Spectral element {FNPF} simulation of focused wave groups impacting a fixed {FPSO}-type body",
author = "Engsig-Karup, \{Allan P.\} and Claes Eskilsson",
year = "2019",
volume = "29",
pages = "141--148",
journal = "International Journal of Offshore and Polar Engineering",
publisher = "International Society of Offshore \& Polar Engineers",
number = "2",
}

@article{engsigkarup2021efficient,
    title = {{An efficient p-multigrid spectral element model for fully nonlinear water waves and fixed bodies}},
    year = {2021},
    journal = {International Journal for Numerical Methods in Fluids},
    author = {Engsig-Karup, Allan P and Laskowski, Wojciech L},
    number = {9},
    pages = {2823--2841},
    volume = {93},
}

@inproceedings{engsigkarup2013fast,
	Author = {Engsig-Karup, A. P. and Glimberg, L. S. and Nielsen, A. S. and Lindberg, O.},
	Booktitle = {Designing Scientific Applications on GPUs},
	Editor = {Raph{\"a}el Couturier},
	Publisher = {CRC Press / Taylor \& Francis Group},
	Series = {Lecture notes in computational science and engineering},
	Title = {Fast hydrodynamics on heterogenous many-core hardware},
	Year = {2013}}

@article{engsigkarup2019mixed,
    title = {A Mixed {E}ulerian–{L}agrangian Spectral Element Method for Nonlinear Wave Interaction with Fixed Structures},
    year = {2019},
    journal = {Water Waves},
    author = {Engsig-Karup, Allan P and Monteserin, Carlos and Eskilsson, Claes},
    number = {2},
    pages = {315--342},
    volume = {1},
}

@article{engsigkarup2016stabilised,
    title = {{A stabilised nodal spectral element method for fully nonlinear water waves}},
    year = {2016},
    journal = {Journal of Computational Physics},
    author = {Engsig-Karup, A P and Eskilsson, C and Bigoni, D},
    pages = {1--21},
    volume = {318},
}

@article{engsigkarup2009efficient,
    title = {{An efficient flexible-order model for 3{D} nonlinear water waves}},
    year = {2009},
    journal = {Journal of Computational Physics},
    author = {Engsig-Karup, A P and Bingham, H B and Lindberg, O},
    number = {6},
    pages = {2100--2118},
    volume = {228},
}

@article{fenton1988numerical,
title = {The numerical solution of steady water wave problems},
journal = {Computers \& Geosciences},
volume = {14},
number = {3},
pages = {357-368},
year = {1988},
author = {J. Fenton},
}

@article{fedkiw1999non,
  title={A non-oscillatory {E}ulerian approach to interfaces in multimaterial flows (the ghost fluid method)},
  author={Fedkiw, Ronald P and Aslam, Tariq and Merriman, Barry and Osher, Stanley},
  journal={Journal of computational physics},
  volume={152},
  number={2},
  pages={457--492},
  year={1999},
  publisher={Elsevier}
}

@article{hirt1981volume,
  title={{Volume of fluid (VOF) method for the dynamics of free boundaries}},
  author={Hirt, Cyril W and Nichols, Billy D},
  journal={Journal of Computational Physics},
  volume={39},
  number={1},
  pages={201--225},
  year={1981},
  publisher={Elsevier}
}

@article{hanssen2021potential,
  title={A potential flow method combining immersed boundaries and overlapping grids: Formulation, validation and verification},
  author={Hanssen, Finn-Christian Wickmann and Greco, Marilena},
  journal={Ocean Engineering},
  volume={227},
  pages={108841},
  year={2021},
  publisher={Elsevier}
}

@phdthesis{hicks2020development,
title = "Development of a high-order potential flow solver for nonlinear wave-structure interaction",
author = "Hicks, {Jacob Bjarke Hansen}",
year = "2020",
language = "English",
series = "DCAMM Special Report",
number = "S284",
publisher = "Technical University of Denmark",
school = "Technical University of Denmark",
}

@article{hicks2021nonlinear,
  title={Nonlinear wave generation using a heaving wedge},
  author={Hicks, Jacob BH and Bingham, Harry B and Read, Robert W and Engsig-Karup, Allan P},
  journal={Applied Ocean Research},
  volume={108},
  pages={102540},
  year={2021},
  publisher={Elsevier}
}

@book{hesthaven2008nodal,
  title={Nodal discontinuous Galerkin methods: algorithms, analysis, and applications},
  author={Hesthaven, Jan S and Warburton, Tim},
  year={2008},
  publisher={Springer}
}

@article{kreiss1972comparison,
  title={Comparison of accurate methods for the integration of hyperbolic equations},
  author={Kreiss, Heinz-Otto and Oliger, Joseph},
  journal={Tellus},
  volume={24},
  number={3},
  pages={199--215},
  year={1972},
  publisher={Taylor \& Francis}
}

@article{knoblauch2026development,
  title={Development of New Free Surface Methods for a Fixed-Grid Fully Nonlinear Potential Flow Solver Using the Immersed Boundary Method},
  author={Knoblauch, Fabian and Wang, Widar and Bihs, Hans},
  journal={International Journal for Numerical Methods in Fluids},
  year={2026},
  publisher={Wiley Online Library}
}

@book{karniadakis2005spectral,
    title = {{Spectral/hp element methods for computational fluid dynamics}},
    year = {2005},
    author = {Karniadakis, George and Sherwin, Spencer},
    publisher = {OUP Oxford}
}

@article{klahn2021new,
  title={A new $\sigma$-transform based {F}ourier-{L}egendre-{G}alerkin model for nonlinear water waves},
  author={Klahn, Mathias and Madsen, Per A and Fuhrman, David R},
  journal={International Journal for Numerical Methods in Fluids},
  volume={93},
  number={1},
  pages={220--248},
  year={2021},
  publisher={Wiley Online Library}
}

@phdthesis{kontos2016robust,
title = "Robust numerical methods for nonlinear wave-structure interaction in a moving frame of reference",
author = "Stavros Kontos",
year = "2016",
language = "English",
series = "DCAMM Special Report",
number = "S217",
publisher = "Technical University of Denmark",
school = "Technical University of Denmark",
}

@article{longuet1976deformation,
  title={The deformation of steep surface waves on water-{I}. {A} numerical method of computation},
  author={Longuet-Higgins, Michael Selwyn and Cokelet, ED0346},
  journal={Proceedings of the Royal Society of London. A. Mathematical and Physical Sciences},
  volume={350},
  number={1660},
  pages={1--26},
  year={1976},
  publisher={The Royal Society London}
}

@article{larsen1983open,
title = {Open boundaries in short wave simulations — A new approach},
journal = {Coastal Engineering},
volume = {7},
number = {3},
pages = {285-297},
year = {1983},
author = {Jesper Larsen and Henry Dancy}
}

@article{li1997three,
  title={A three dimensional multigrid model for fully nonlinear water waves},
  author={Li, Bin and Fleming, Christopher A},
  journal={Coastal Engineering},
  volume={30},
  number={3-4},
  pages={235--258},
  year={1997},
  publisher={Elsevier}
}

@article{maxworthy1976experiments,
  title={Experiments on collisions between solitary waves},
  author={Maxworthy, T},
  journal={Journal of Fluid Mechanics},
  volume={76},
  number={1},
  pages={177--186},
  year={1976},
  publisher={Cambridge University Press}
}

@article{madsen2002new,
  title={A new {B}oussinesq method for fully nonlinear waves from shallow to deep water},
  author={Madsen, Per A and Bingham, Harry B and Liu, Hua},
  journal={Journal of Fluid Mechanics},
  volume={462},
  pages={1--30},
  year={2002},
  publisher={Cambridge University Press}
}

@article{main2018shifted_part1,
title = {The shifted boundary method for embedded domain computations. {P}art {I}: {P}oisson and {S}tokes problems},
journal = {Journal of Computational Physics},
volume = {372},
pages = {972-995},
year = {2018},
author = {A. Main and G. Scovazzi},
}

@article{main2018shifted_part2,
title = {The shifted boundary method for embedded domain computations. {P}art {II}: Linear advection–diffusion and incompressible {N}avier–{S}tokes equations},
journal = {Journal of Computational Physics},
volume = {372},
pages = {996-1026},
year = {2018},
author = {A. Main and G. Scovazzi},
}

@article{modderman2025application,
  title={Application of unfitted finite element methods for estimating added mass and added damping in floating structures},
  author={Modderman, Jan and Colom{\'e}s, Oriol},
  journal={Advances in Computational Science and Engineering},
  volume={4},
  pages={142--167},
  year={2025},
  publisher={Advances in Computational Science and Engineering}
}

@phdthesis{modderman2026unfitted,
title = "Unfitted finite element methods for floating structure hydrodynamics",
author = "J. Modderman",
year = "2026",
school = "Delft University of Technology",
}

@article{naga2005polynomial,
  title={The polynomial-preserving recovery for higher order finite element methods in 2{D} and 3{D}},
  author={Naga, Ahmed},
  journal={Discrete \& Continuous Dynamical Systems-B},
  year={2005}
}

@article{nouveau2019high,
  title={High-order gradients with the shifted boundary method: an embedded enriched mixed formulation for elliptic {PDE}s},
  author={Nouveau, Leo and Ricchiuto, Mario and Scovazzi, Guglielmo},
  journal={Journal of Computational Physics},
  volume={398},
  pages={108898},
  year={2019},
  publisher={Elsevier}
}

@book{newman2018marine,
  title={Marine hydrodynamics},
  author={Newman, John Nicholas},
  year={2018},
  publisher={MIT press}
}

@phdthesis{nielsen2020computational,
title = "Computational methods for wave-structure interaction: Numerical analysis of a RBF-based method",
author = "Nielsen, \{Morten Eggert\}",
year = "2020",
publisher = "Aalborg Universitetsforlag",
school = "Aalborg University",
address = "Denmark",
}

@article{patera1984spectral,
title = {A spectral element method for fluid dynamics: Laminar flow in a channel expansion},
journal = {Journal of Computational Physics},
volume = {54},
number = {3},
pages = {468-488},
year = {1984},
author = {Anthony T Patera},
}

@article{peskin1972flow,
title = {Flow patterns around heart valves: A numerical method},
journal = {Journal of Computational Physics},
volume = {10},
number = {2},
pages = {252-271},
year = {1972},
author = {Charles S Peskin},
}

@article{rim2026naturally,
title = {A naturally sharpened level-set formulation for incompressible free-surface flows},
journal = {Computer Methods in Applied Mechanics and Engineering},
volume = {453},
pages = {118798},
year = {2026},
author = {Jongmin Rim and Jinhui Yan and Yuri Bazilevs},
}

@article{ransley2019blind,
  title={A blind comparative study of focused wave interactions with a fixed {FPSO}-like structure ({CCP-WSI} {B}lind {T}est {S}eries 1)},
  author={Ransley, Edward and Yan, Shiqiang and Brown, Scott Andrew and Mai, Tri and Graham, David and Ma, Qingwei and Musiedlak, Pierre-Henri and Engsig-Karup, Allan Peter and Eskilsson, Claes and Li, Qian and others},
  journal={International Journal of Offshore and Polar Engineering},
  volume={29},
  number={02},
  pages={113--127},
  year={2019},
  publisher={ISOPE}
}

@article{robertson1999free,
    title = {Free-Surface Flow Simulation Using hp/Spectral Elements},
    year = {1999},
    journal = {Journal of Computational Physics},
    author = {Robertson, Iain and Sherwin, Spencer},
    number = {1},
    pages = {26--53},
    volume = {155}
}

@article{shao2014harmonic,
title = {A harmonic polynomial cell ({HPC}) method for 3{D} {L}aplace equation with application in marine hydrodynamics},
journal = {Journal of Computational Physics},
volume = {274},
pages = {312-332},
year = {2014},
author = {Yan-Lin Shao and Odd M. Faltinsen},
}

@article{song2018shifted,
title = {The shifted boundary method for hyperbolic systems: Embedded domain computations of linear waves and shallow water flows},
journal = {Journal of Computational Physics},
volume = {369},
pages = {45-79},
year = {2018},
author = {T. Song and A. Main and G. Scovazzi and M. Ricchiuto},
}

@article{savitzky1964smoothing,
  title={Smoothing and differentiation of data by simplified least squares procedures.},
  author={Savitzky, Abraham and Golay, Marcel JE},
  journal={Analytical Chemistry},
  volume={36},
  number={8},
  pages={1627--1639},
  year={1964},
  publisher={ACS Publications}
}

@article{shankar2018hyperviscosity,
  title={Hyperviscosity-based stabilization for radial basis function-finite difference ({RBF-FD}) discretizations of advection--diffusion equations},
  author={Shankar, Varun and Fogelson, Aaron L},
  journal={Journal of Computational Physics},
  volume={372},
  pages={616--639},
  year={2018},
  publisher={Elsevier}
}

@article{tsai1996computation,
  title={Computation of nonlinear free-surface flows},
  author={Tsai, Wu-Ting and Yue, Dick KP},
  journal={Annual Review of Fluid Mechanics},
  volume={28},
  number={1},
  pages={249--278},
  year={1996},
  publisher={Annual Reviews 4139 El Camino Way, PO Box 10139, Palo Alto, CA 94303-0139, USA}
}

@article{tong2024adaptive,
  title={An adaptive harmonic polynomial cell method for three-dimensional fully nonlinear wave-structure interaction with immersed boundaries},
  author={Tong, Chao and Shao, Yanlin and Bingham, Harry B and Hanssen, Finn-Christian W},
  journal={Physics of Fluids},
  volume={36},
  number={3},
  year={2024},
  publisher={AIP Publishing}
}

@article{visbech2024solving,
  title={Solving the complete pseudo-impulsive radiation and diffraction problem using a spectral element method},
  author={Visbech, Jens and Engsig-Karup, Allan P and Bingham, Harry B},
  journal={Computer Methods in Applied Mechanics and Engineering},
  volume={423},
  pages={116871},
  year={2024},
  publisher={Elsevier}
}

@inproceedings{visbech2024high,
  title={A high-order shifted boundary method for water waves and floating bodies},
  author={Visbech, Jens and Engsig-Karup, Allan Peter and Bingham, Harry B and Amini-Afshar, Mostafa and Ricchiuto, Mario},
  booktitle={Proceedings: The 39st International Workshop on Water Waves and Floating Bodies, Sct. Andrews, Scotland},
  year={2024}
}

@article{visbech2025spectral,
  title={A spectral element solution of the {P}oisson equation with shifted boundary polynomial corrections: influence of the surrogate to true boundary mapping and an asymptotically preserving {R}obin formulation},
  author={Visbech, Jens and Engsig-Karup, Allan P and Ricchiuto, Mario},
  journal={Journal of Scientific Computing},
  volume={102},
  number={1},
  pages={11},
  year={2025},
  publisher={Springer}
}

@inproceedings{visbech2026recent,
  title={Recent progress on modeling nonlinear wave propagation and wave-structure interaction using a high-order shifted boundary method: Capabilities, challenges, and perspectives},
  author={Visbech, Jens and Engsig-Karup, Allan Peter and Bingham, Harry B and Ricchiuto, Mario},
  booktitle={Proceedings: The 41st International Workshop on Water Waves and Floating Bodies, Plitvice, Croatia},
  year={2026},
}

@article{visbech2025fnpf,
  title={{FNPF-SEM}: A parallel spectral element model in {F}iredrake for fully nonlinear water wave simulations},
  author={Visbech, Jens and Melander, Anders and Engsig-Karup, Allan Peter},
  journal={The International Journal of High Performance Computing Applications},
  pages={10943420261462394},
  year={2025},
  publisher={SAGE Publications Sage UK: London, England}
}

@inproceedings{wang2013study,
  title={Study on interaction between a solitary wave and a submerged semi-circular cylinder using acceleration potential},
  author={Wang, Lixian and Tang, Hui and Wu, Yanhua},
  booktitle={Proceedings: 23rd International Ocean and Polar Engineering Conference, Anchorage, Alaska, USA},
  year={2013},
}

@article{xu2023high,
title = {A high-order finite difference method with immersed-boundary treatment for fully-nonlinear wave–structure interaction},
journal = {Applied Ocean Research},
volume = {134},
pages = {103535},
year = {2023},
author = {Yan Xu and Harry B. Bingham and Yanlin Shao},
}

@article{xu2021finite,
author = {Xu, Yan and Bingham, Harry B. and Shao, Yanlin},
title = {Finite difference solutions for nonlinear water waves using an immersed boundary method},
journal = {International Journal for Numerical Methods in Fluids},
volume = {93},
number = {4},
pages = {1143-1162},
year = {2021}
}

@article{xu2025weighted,
  title={A weighted shifted boundary method for the {N}avier-{S}tokes equations with immersed moving boundaries},
  author={Xu, Danjie and Colom{\'e}s, Oriol and Main, Alex and Li, Kangan and Atallah, Nabil M and Abboud, Nabil and Scovazzi, Guglielmo},
  journal={Journal of Computational Physics},
  pages={114571},
  year={2025},
  publisher={Elsevier}
}

@article{xu2018spectral,
    title = {{Spectral/hp element methods: Recent developments, applications, and perspectives}},
    year = {2018},
    journal = {Journal of Hydrodynamics},
    author = {Xu, Hui and Cantwell, Chris and Monteserin, Carlos and Eskilsson, Claes and Engsig-Karup, Allan Peter and Sherwin, Spencer},
    month = {5},
    pages = {},
    volume = {30},
}

@article{yang2025octree,
  title={Octree-based adaptive mesh refinement and the shifted boundary method for efficient fluid dynamics simulations},
  author={Yang, Cheng-Hau and Scovazzi, Guglielmo and Krishnamurthy, Adarsh and Ganapathysubramanian, Baskar},
  journal={Advances in Computational Science and Engineering},
  volume={4},
  year={2025},
  publisher={American Institute of Mathematical Sciences}
}

@article{zhou2024solving,
  title={Solving for hydroelastic ship response using a high-order finite difference method on overlapping grids at zero speed},
  author={Zhou, Baoshun and Amini-Afshar, Mostafa and Bingham, Harry B and Shao, Yanlin and Malenica, {\v{S}}ime and Andersen, Matilde H},
  journal={Marine Structures},
  volume={95},
  pages={103602},
  year={2024},
  publisher={Elsevier}
}

% When submitting to arXiv (example) - remember to select all .bib etc. files.:



% \end{thebibliography}

\newpage

\appendix

\section{Comparing effect of stabilization parameters}\label{app:filter_runs}

\begin{table}[H]
\centering
\caption{Testing various stabilization parameters for different waves and resolution with $m = \{1,\ldots,5\}$, $\gamma = \{0,0.1, 1, 10\}$, $kh  = \{1, \pi, 2\pi\}$, $N_x = \{4,8,16\}$, $N_z=\{1,2,4\}$, and $p = \{4,5,6\}$ for $\varepsilon/\varepsilon_{\max} = 90\%$ nonlinearity after $50T$ for analysis in Section \ref{sec:long_time_periodic}. Showing the 50 best runs sorted after $e_{\text{mix}} = e_{\text{dif}}+e_{\text{dis}}$.}
\vspace{1mm}
\begin{tabular}{lllllllll}
\hline
$kh$ & $N_x$ & $N_z$ & $p$& $m$ & $\gamma$ & $e_{\text{dif}}$ & $e_{\text{dis}}$ & $e_{\text{mix}}$ \\ \hline
1             & 16             & 4              & 6            & 4            & 10                & 0.1802                    & 0.0100                    & 0.1902                    \\
1             & 16             & 4              & 6            & 3            & 1                 & 0.1403                    & 0.1100                    & 0.2503                    \\
1             & 16             & 4              & 6            & 5            & 10                & 0.0809                    & 0.2400                    & 0.3209                    \\
1             & 16             & 4              & 5            & 5            & 10                & 0.3970                    & 0.1200                    & 0.5170                    \\
1             & 16             & 2              & 6            & 4            & 10                & 0.1367                    & 0.4501                    & 0.5868                    \\
1             & 16             & 2              & 6            & 3            & 1                 & 0.1059                    & 0.5201                    & 0.6260                    \\
1             & 16             & 2              & 5            & 3            & 1                 & 0.4374                    & 0.2100                    & 0.6474                    \\
1             & 16             & 2              & 6            & 4            & 1                 & 0.0383                    & 0.6901                    & 0.7283                    \\
1             & 16             & 2              & 6            & 5            & 1                 & 0.0292                    & 0.7101                    & 0.7393                    \\
1             & 16             & 2              & 6            & 5            & 10                & 0.0541                    & 0.6901                    & 0.7442                    \\
1             & 16             & 2              & 5            & 4            & 10                & 0.6209                    & 0.2200                    & 0.8410                    \\
1             & 16             & 2              & 5            & 5            & 10                & 0.2021                    & 0.6801                    & 0.8822                    \\
$\pi$         & 16             & 4              & 6            & 3            & 10                & 0.1407                    & 0.7701                    & 0.9108                    \\
$\pi$         & 16             & 4              & 6            & 3            & 1                 & 0.0057                    & 1.1001                    & 1.1058                    \\
$\pi$         & 16             & 4              & 6            & 4            & 10                & 0.0100                    & 1.1001                    & 1.1101                    \\
$\pi$         & 16             & 4              & 6            & 4            & 1                 & 0.0197                    & 1.1301                    & 1.1498                    \\
$\pi$         & 16             & 4              & 6            & 5            & 10                & 0.0213                    & 1.1301                    & 1.1514                    \\
1             & 16             & 4              & 5            & 3            & 1                 & 0.6113                    & 0.5801                    & 1.1913                    \\
$\pi$         & 16             & 4              & 6            & 5            & 1                 & 0.0682                    & 1.2401                    & 1.3083                    \\
1             & 16             & 2              & 5            & 4            & 1                 & 0.3550                    & 0.9901                    & 1.3451                    \\
$\pi$         & 8              & 4              & 6            & 3            & 1                 & 0.7986                    & 0.5801                    & 1.3787                    \\
$\pi$         & 16             & 4              & 5            & 3            & 10                & 0.5505                    & 0.8801                    & 1.4306                    \\
$\pi$         & 8              & 2              & 6            & 4            & 10                & 1.4482                    & 0.2800                    & 1.7282                    \\
$\pi$         & 16             & 4              & 6            & 2            & 1                 & 0.1148                    & 1.6802                    & 1.7950                    \\
1             & 16             & 4              & 5            & 4            & 10                & 0.7869                    & 1.0401                    & 1.8270                    \\
$2\pi$        & 8              & 4              & 6            & 4            & 10                & 1.2064                    & 0.7001                    & 1.9065                    \\
$\pi$         & 16             & 4              & 5            & 4            & 10                & 0.0195                    & 1.9602                    & 1.9797                    \\
$\pi$         & 16             & 4              & 5            & 3            & 1                 & 0.0229                    & 1.9602                    & 1.9831                    \\
$\pi$         & 8              & 4              & 6            & 4            & 1                 & 0.5827                    & 1.4701                    & 2.0528                    \\
$\pi$         & 16             & 4              & 5            & 5            & 10                & 0.0262                    & 2.0502                    & 2.0764                    \\
1             & 16             & 2              & 4            & 3            & 1                 & 1.6516                    & 0.4601                    & 2.1117                    \\
$\pi$         & 16             & 4              & 5            & 4            & 1                 & 0.0338                    & 2.0802                    & 2.1140                    \\
1             & 16             & 2              & 6            & 3            & 10                & 0.9168                    & 1.3001                    & 2.2169                    \\
$\pi$         & 8              & 2              & 6            & 3            & 1                 & 0.5710                    & 1.7002                    & 2.2712                    \\
1             & 16             & 4              & 6            & 3            & 10                & 0.8190                    & 1.4601                    & 2.2791                    \\
$2\pi$        & 8              & 4              & 6            & 3            & 1                 & 0.3104                    & 2.0402                    & 2.3506                    \\
1             & 16             & 2              & 4            & 5            & 10                & 1.0841                    & 1.2901                    & 2.3742                    \\
$\pi$         & 16             & 4              & 5            & 2            & 1                 & 0.1662                    & 2.2902                    & 2.4564                    \\
1             & 16             & 2              & 5            & 2            & 1                 & 0.3757                    & 2.2002                    & 2.5759                    \\
$\pi$         & 8              & 2              & 6            & 4            & 1                 & 0.0920                    & 2.7003                    & 2.7923                    \\
$\pi$         & 8              & 2              & 6            & 5            & 10                & 0.2590                    & 2.8203                    & 3.0793                    \\
$\pi$         & 8              & 2              & 5            & 3            & 1                 & 1.3131                    & 2.0102                    & 3.3233                    \\
$\pi$         & 16             & 4              & 4            & 2            & 1                 & 1.7907                    & 1.5702                    & 3.3609                    \\
$2\pi$        & 8              & 4              & 6            & 5            & 10                & 0.4335                    & 2.9403                    & 3.3738                    \\
$2\pi$        & 8              & 4              & 6            & 4            & 1                 & 0.3912                    & 2.9903                    & 3.3815                    \\
$\pi$         & 16             & 2              & 6            & 3            & 10                & 0.3251                    & 3.0703                    & 3.3954                    \\
$\pi$         & 8              & 2              & 5            & 5            & 10                & 0.5048                    & 2.9403                    & 3.4451                    \\
$\pi$         & 16             & 4              & 4            & 4            & 10                & 0.4282                    & 3.0903                    & 3.5185                    \\
$2\pi$        & 8              & 4              & 5            & 3            & 1                 & 0.8164                    & 2.7603                    & 3.5767                    \\
$\pi$         & 16             & 4              & 4            & 3            & 1                 & 0.3869                    & 3.2303                    & 3.6172                    \\ \hline
\end{tabular}
\label{tab:best_runs_with_different_filters}
\end{table}

\newpage

\section{Free surface wave elevations for the submerged bar test}\label{app:bartest}

\begin{figure}[h]
    \centering
    \includegraphics{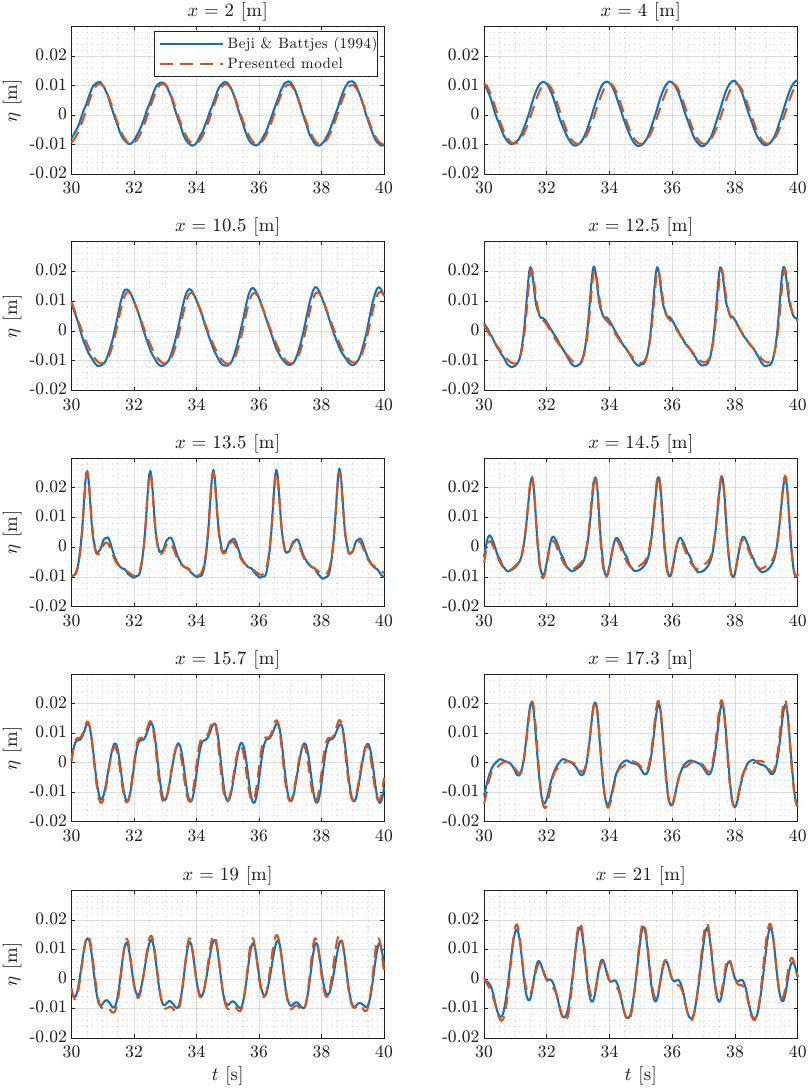}
    \caption{Time series of the free surface wave elevation at 10 different wave gauge locations for the submerged bar test in Section \ref{sec:bartest}.}
    \label{fig:bar_test_time_series}
\end{figure}

\newpage

\section{Unfitted mesh for the semi-circular bathymetry bump}\label{app:mesh_semi_circular_bump}

\begin{figure}[h]
    \centering
    \includegraphics[scale=1]{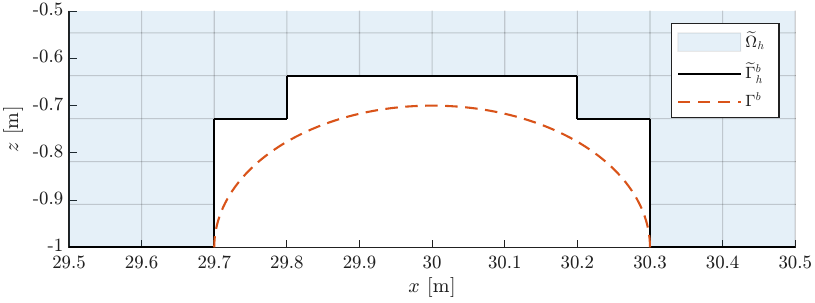}
    \caption{Zoomed in around the unfitted mesh of the semi-circular bathymetry bump.}
    \label{fig:unfitted_semi_circular_bump}
\end{figure}

\end{document}